%% file: ex_article.tex
\documentclass[preprint,hidelinks,onefignum,onetabnum]{siamart220329}

\input{ex_shared}

\ifpdf
\hypersetup{
  pdftitle={An Example Article},
  pdfauthor={}
}
\fi

\acrodef{FAS}{full approximation scheme}
\acrodef{DSL}{domain specific language}
\acrodef{RHS}{right-hand side}
\acrodef{GP}{Genetic programming}
\acrodef{G3P}{Grammar-guided genetic programming}
\acrodef{CFG}{context-free grammar}
\acrodef{PDE}{partial differential equation}
\acrodef{EA}{Evolutionary Algorithm}
\acrodef{AI}{artificial intelligence}
\acrodef{AMG}{algebraic multigrid}
\acrodef{GMG}{geometric multigrid}
\acrodef{CG}{conjugate gradient}
\acrodef{PCG}{preconditioned conjugate gradient}
\acrodef{GS}{Gauss-Seidel}
\acrodef{AMG-PCG}{AMG preconditioned CG}
\acrodef{CGC}{coarse-grid correction}

\begin{document}

\maketitle

\begin{abstract}
\input{siamart_220329/contents/abstract}
\end{abstract}



\input{siamart_220329/contents/introduction}
\input{siamart_220329/contents/related_work}
\input{siamart_220329/contents/background}

\input{siamart_220329/contents/method}
\input{siamart_220329/contents/experiments}

\input{siamart_220329/contents/results}

\input{siamart_220329/contents/conclusion}

\input{siamart_220329/contents/acknowledgements}

\bibliographystyle{siamplain}
\bibliography{references}
\newpage
\input{siamart_220329/contents/appendix}
\end{document}

%% file: ex_shared.tex
\usepackage{lipsum}
\usepackage{amsfonts}
\usepackage{graphicx}
\usepackage{epstopdf}

\usepackage[most]{tcolorbox}

\usepackage{xcolor}
\colorlet{opcol}{blue!70!black}
\colorlet{ntcol}{black!60}
\colorlet{parcol}{green!45!black}

\usepackage[colorinlistoftodos,prependcaption,textsize=small]{todonotes}
\newcommand{\todURi}[1] {\todo[color=green!25,inline,size=\footnotesize]{{\bf U:} #1}}
\newcommand{\todUR}[1]%
{\todo[color=green!25,linecolor=darkgray,size=\footnotesize]{{\bf U:} #1} 
}

\usepackage{algorithmic}
\ifpdf
  \DeclareGraphicsExtensions{.eps,.pdf,.png,.jpg}
\else
  \DeclareGraphicsExtensions{.eps}
\fi

\newsiamremark{remark}{Remark}
\newsiamremark{hypothesis}{Hypothesis}
\crefname{hypothesis}{Hypothesis}{Hypotheses}
\newsiamthm{claim}{Claim}

\headers{Grammar-based AMG With Evolutionary Algorithms}{D. Parthasarathy}

\title{Automated Grammar-based Algebraic Multigrid Design With Evolutionary Algorithms}

\author{
  Dinesh Parthasarathy%
  \thanks{Friedrich-Alexander-Universität Erlangen-Nürnberg (FAU), Erlangen, Germany
    (\email{dinesh.parthasarathy@fau.de},
     \email{harald.koestler@fau.de},
     \email{ulrich.ruede@fau.de}).}
  \and
  Wayne Mitchell%
  \thanks{Lawrence Livermore National Laboratory, USA
    (\email{mitchell82@llnl.gov},
     \email{arjun@llnl.gov}).}
  \and
  Arjun Gambhir\footnotemark[2]
  \and
  Harald Köstler\footnotemark[1]
  \and
  Ulrich Rüde\footnotemark[1] \thanks{CERFACS, Toulouse, France and Department of Applied Mathematics, VSB-Technical University of Ostrava, Ostrava, Czech Republic}
}

\usepackage{amsopn}


%% file: siamart_220329/contents/abstract.tex
Although multigrid is asymptotically optimal for solving many important partial differential equations, its efficiency relies heavily on the careful selection of the individual algorithmic components. In contrast to recent approaches that can optimize certain multigrid components using deep learning techniques, we adopt a complementary strategy, employing evolutionary algorithms to construct efficient multigrid cycles from proven algorithmic building blocks. Here, we will present its application to generate efficient algebraic multigrid methods with so-called \emph{flexible cycling}, that is, level-specific smoothing sequences and non-recursive cycling patterns. The search space with such non-standard cycles is intractable to navigate manually, and is generated using genetic programming (GP) guided by context-free grammars. Numerical experiments with the linear algebra library, \emph{hypre}, demonstrate the potential of these non-standard GP cycles to improve multigrid performance both as a solver and a preconditioner.

%% file: siamart_220329/contents/introduction.tex
\section{Introduction}
Solving partial differential equations (PDEs) numerically on modern computers has become ubiquitous in science and engineering. However, developing scientific codes for this purpose demands a highly interdisciplinary effort, requiring experts from physics, numerical analysis, and high-performance computing. Over the past several decades, significant progress in developing such codes has produced a rich software ecosystem for solving PDEs. 
In this article we will discuss tools to synthesize such codes  in the framework of time-dependent PDEs of the form
\begin{equation}
    u_t + {N}(u)(\Vec{x},t) = f(\vec{x},t) \quad \text{in } \Omega \times [0,T],
    \label{eq:continuousform}
\end{equation}
where ${N}$ is a spatial derivative operator.
\emph{Domain-specific languages} allow such systems to be expressed in a high-level \emph{pen-and-paper} form, and can be automatically translated into efficient implementations \cite{alns2014unified-a44,bauer2019code-f58,rathgeber2016firedrake-347,Lengauer2020}. Furthermore, \emph{multi-physics frameworks} allow users to tackle complex, coupled systems of PDEs, providing a variety of space and time discretization choices \cite{DUNE_Bastian2021, MFEM_Anderson2021,DealII_Bangerth2007,waLBerla_Bauer2021}. For example,  a semi-discrete form of \cref{eq:continuousform} could be
\begin{equation}
{M_h} \, \dot{U}(t) + {N}_h(U)(t) = {F_h}(t) \quad \text{in } \Omega_h \times [0,T],
\label{eq:semidiscrete}
\end{equation}
where ${M_h}$ is the mass matrix and ${N}_h$ represents the 
discretized differential operator. Considering a special case of linear PDEs, $N_h(U) = L_hU$, with an implicit backward Euler time integrator, each time step involves solving a large sparse linear system

$$
\underbrace{\left({M_h} + \Delta t {L}_h \right)}_{{A}} \underbrace{{U}^{(n+1)}}_\mathbf{x} = \underbrace{{M_h} {U}^{(n)} + \Delta t {F_h}}_{\mathbf{b}},
$$
of the form 
\begin{equation}
\label{eq:linear}
    {A} \mathbf{x} = \mathbf{b}.
\end{equation}
In practice the solution is often offloaded to a set of specialized \emph{solver libraries}, such as those in \cite{heroux2012new-102, falgout2006numerical-576, MUMPS:2, mills2021}.  
Consequently, designing efficient linear solvers is crucial, as these kernels typically dominate the overall execution time of simulation codes.
Although the PDE software stack is often designed around the principle of separation of concerns, spanning physical modeling, algorithm design, and parallelization techniques, in practice these aspects remain deeply intertwined.
Superior performance can, in fact, often only be achieved with an integrated co-design of application and solver \cite{kohl2024fundamental, bohm2025code}.
However, it is challenging for application users 
to design efficient, application-specific solvers without a multi-disciplinary background.
\Ac{AMG} methods have become one of the standard workhorses for solving systems as in \cref{eq:linear}: they can achieve linear complexity with respect to the number of unknowns, and are therefore well suited for large-scale simulations. We provide a concise overview of AMG in \cref{sec:AMG}. However, an effective method (i) exploits the algebraic properties of the problem, such as the matrix sparsity pattern, condition number, strength of connections between unknowns; (ii) has good stability and convergence properties with the chosen numerical precision; and (iii) is easy to parallelize and scale to large problems. Hence, designing efficient AMG methods for a specific problem requires expert knowledge and can be time-intensive. We address the scope of designing efficient AMG methods automatically, tailored to the problem at hand, so that non-specialists can deploy them easily within their simulation codes.

\begin{figure}
    \centering
    \includegraphics[width=1\linewidth]{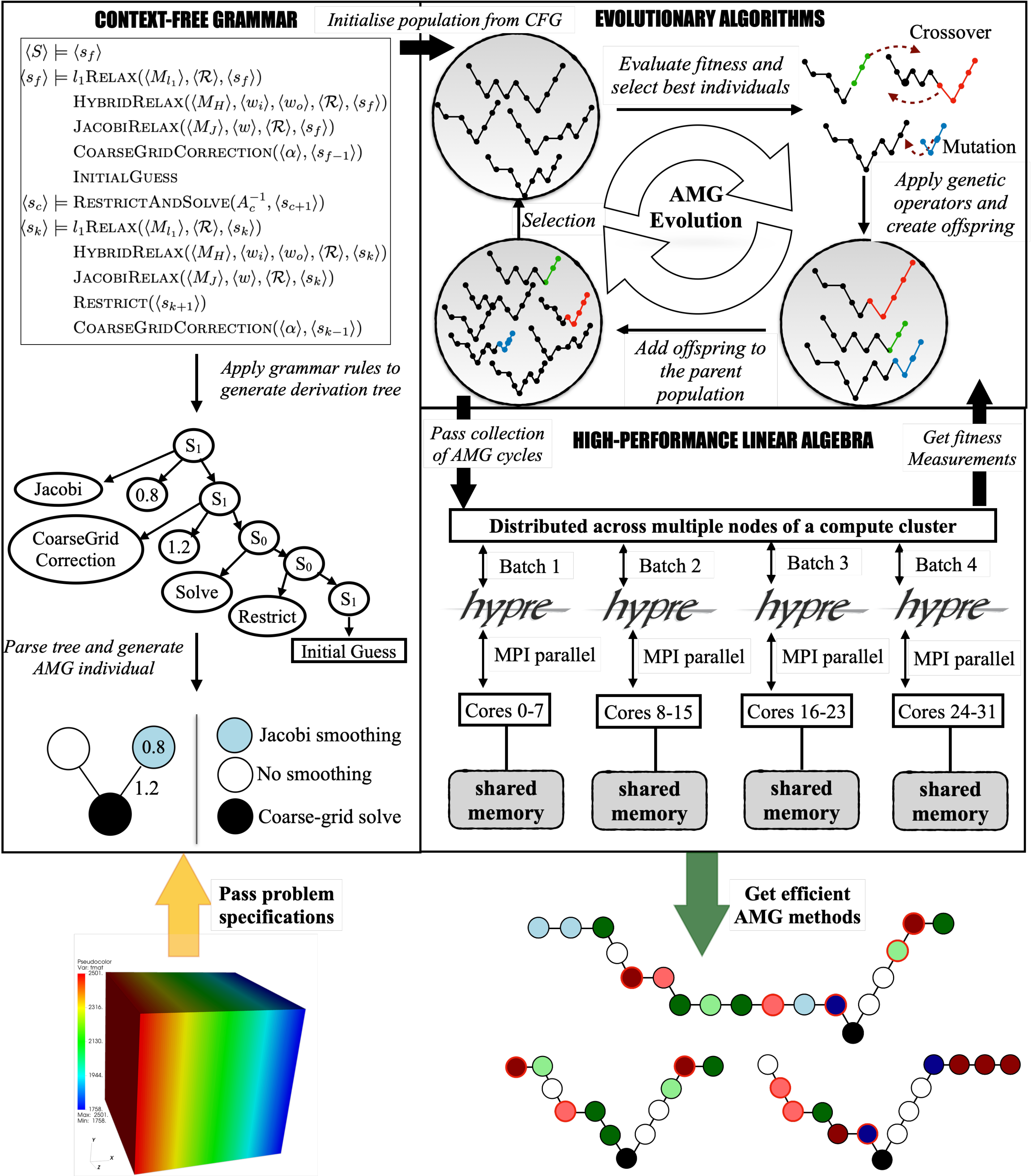}
    \caption{A high-level overview of the proposed automated AMG design workflow: given an input problem, the approach uses grammars, evolutionary algorithms, and high-performance solvers to generate flexible AMG cycles. Nodes in the resulting cycles are color-coded to indicate different smoothing operations. }
    \label{fig:introflowchart}
\end{figure}

As a possible solution, we propose an automated approach, using context-free grammars (CFGs) that encode the syntactic and algorithmic constraints of AMG methods, and guide an evolutionary algorithm to design solvers for a specific problem. This is cast as a program-synthesis task using grammar-guided genetic programming (GP) (see \cref{sec:g3p} for a short primer), where AMG programs are generated not only by tuning parameters but also by altering the algorithmic scheme within the limits imposed by the grammar. In particular, we formulate a grammar that allows i) \emph{arbitrary} and \emph{non-recursive cycling}, ii) \emph{use of different relaxation schemes} at each step along the cycle. This approach relies solely on the solution time and convergence rates of the candidate solvers as fitness without additional gradient information, and thus, unlike deep learning approaches, can be integrated non-intrusively with existing solver frameworks like \emph{hypre} (\cref{fig:introflowchart}).

We apply this approach to design efficient multigrid cycles for a given AMG setup in \emph{hypre}, i) as a solver in a stationary 3D anisotropic Poisson problem (\cref{sec:amgsolver}), and ii) as a preconditioner for conjugate gradient (CG), in a time-dependent radiation-diffusion/electron-conduction benchmark problem (\cref{sec:amgpredconditioner}). The GP-designed AMG methods outperform both the default and a tuned AMG. The GP-generated methods also generalize well across problem sizes (\cref{subsubsec:evalweakscaling}) and changes in the system matrix during time-stepping (\cref{subsubsec:gentimesteps}). We also show that our generated AMG methods remain compatible with other \emph{hypre} interfaces; more specifically, we test \emph{hypre's} hybrid solver interface\footnote{https://hypre.readthedocs.io/en/latest/solvers-hybrid.html} that switches adaptively during the simulation from a cheap diagonal preconditioner to a more expensive AMG preconditioner, based on the solver convergence at the current iterate (\cref{sec:hybridsolvers}).

The main contributions of this work are:
\begin{itemize}

    \item An evolutionary approach for automated AMG design that integrates non-intrusively with existing linear solver libraries like \emph{hypre}, where programs are constructed in a bottom-up approach, hard-constrained by AMG rules defined in a CFG, and therefore automatically respecting algorithmic integrity and software compatibility.
    \item Discovery of novel, flexible cycling strategies that outperform default and tuned AMG configurations.
    \item Demonstration of good weak scalability and robustness to system matrix perturbations for the GP-designed flexible AMG methods.
\end{itemize}

\color{blue}

\color{black}

%% file: siamart_220329/contents/related_work.tex
Previous works have explored using deep learning to learn individual multigrid components, including smoothers \cite{huang2022learning-988}, intergrid operators \cite{pmlr-v97-greenfeld19a,10.5555/3524938.3525540, Katrutsa2020, katrutsa2017deep}, coarsening schemes \cite{Taghibakhshi2021OptimizationBasedAM}, and strong threshold parameters \cite{caldana2024deep-353, moore2025graph-f8e}. Other approaches attempt to learn the entire multigrid solver, for instance, by modeling it using U-Nets with learnable parameters \cite{10.5555/3692070.3692760}. We view these works as complementary to ours; while they focus on learning individual multigrid components, our approach constructs efficient AMG cycles from existing and proven algorithmic building blocks. Also, GP operates on a representation of the solver as a \emph{computer program}, agnostic to the underlying technology; whether neural-network-based, mesh-based, or both, and thus can co-exist with and complement other learning approaches. The use of CFGs with GP for multigrid design was first introduced by Schmitt et al. \cite{Schmitt2021,Schmitt2022} for geometric multigrid, and we build on these concepts in our current work for AMG design.

\color{black}

%% file: siamart_220329/contents/background.tex
\section{Background}
\label{sec:background}
\subsection{Algebraic Multigrid (AMG)}
\label{sec:AMG}
Iterative methods are well-suited to solve large sparse linear systems of the form in \cref{eq:linear}. At every iteration $i$, the solution is updated as
\begin{equation}
\label{eq:iterative}
\mathbf{x}^{(i+1)}
=
\underbrace{\left(I - B^{-1} A\right)}_{T}\mathbf{x}^{(i)}
+
B^{-1}\,\mathbf{b},
\end{equation}
where $B$ is an approximation to ${A}$ resulting in an iteration operator $T$, determined by the choice of the underlying method. 
When the exact solution $\mathbf{x^*}$ is unknown, 
the norm of the residual $\mathbf{r}^{(i)} = \mathbf{b} - A \mathbf{x}^{(i)}$ 
can be used as a measure of solution accuracy. The iteration stops when a prescribed stopping criterion, for example,  $\lVert \mathbf{r}^{(i)} \rVert_2 \le \varepsilon$ is fulfilled. 
Multigrid methods are a class of algorithms that apply such iterative schemes on a hierarchy of grids, $L, L-1, \dots 2,1,0$, with $L+1$ levels ordered from the finest to the coarsest. At a given grid level $l$, two complementary operations are applied: i) \emph{smoothing}, or the application of an iteration operator $T=S_l$ that eliminates high-frequency error modes, and ii) \emph{coarse-grid correction}, where $T=C_l$, and involves resolving lower-frequency errors by transferring information to a coarser grid, solving a coarser system, and using this to improve the fine grid solution. 
The multigrid operator is given by 
\begin{equation}
\label{eq:MGoperator}
    M_l = S_l^{\nu_2}\,C_l\,S_l^{\nu_1},
\end{equation} where  $\nu_1,\nu_2$ are the number of \emph{pre-smoothing} and \emph{post-smoothing} steps. For a two-grid method $C_l =I - P_l(A_{l-1})^{-1}R_l\,A_l$, with $P_l, R_l$ denoting the prolongation and restriction operators. In a multi-level method, $(A_{l-1})^{-1}$ is approximated using multigrid $M_{l-1}$, and applied recursively across multiple levels. In AMG, the intergrid operators and the coarse-grid operators $A_{l}$ at all  levels are constructed directly from the linear system during the \emph{setup phase}. The resulting multigrid hierarchy is then employed with a suitable smoother and a cycle type (V-, W-, or F-cycles) in the \emph{solve phase}. The latter is the focus of this paper. The cycle type, for example, determines the order in which information at different grid levels is accessed and exchanged. Standard cycle types are recursive, but in this paper, we explore non-standard ones, referred to as \emph{flexible cycles} (\cref{sec:flexibleAMG}). The efficiency of the AMG solve phase for a given system to a specified residual tolerance may be measured by the i) time to solution $T$, ii) number of AMG iterations, $N$, or iii) the convergence rate, $\rho = \left (\frac{\lVert r^{(N)} \rVert_2}{\lVert r^{(0)} \rVert_2} \right )^{1/N}$. For additional details on AMG concepts please refer to
\cite{Briggs,Stuben2000AMGIntro,falgout2006introduction-70d}. 
\subsection{Parallel smoothers in \emph{hypre}}
When solving linear systems on a parallel computer with \emph{$p$} processors in \emph{hypre}, the matrix equation \cref{eq:linear} is partitioned as
\begin{equation}
\begin{pmatrix}
A_{11} & \cdots & A_{1p} \\
\vdots & \ddots & \vdots \\
A_{p1} & \cdots & A_{pp}
\end{pmatrix}
\begin{pmatrix}
\mathbf{x_1} \\ \vdots \\ \mathbf{x_p}
\end{pmatrix}
=
\begin{pmatrix}
\mathbf{b_1} \\ \vdots \\ \mathbf{b_p}
\end{pmatrix},
\label{eq:block_system}
\end{equation}
where each processor stores a single block row.
Our GP-designed AMG methods choose from several parallel smoothers of the form
$B = \operatorname{diag}(B_1, B_2, \ldots, B_p)$ (as notated in \cref{eq:iterative}). For each block $A_{kk}$,  considering a standard splitting $A_{kk} = D_k - L_k - U_k$, (where $D_k$ is the diagonal and $-L_k,-U_k$ are the lower and upper triangular parts of $A_{kk}$) a parallel \emph{hybrid} Gauss-Seidel forward (GSF) smoother, for example,  is defined as
\begin{equation}\label{eq:parallel_gsf}
B_{\mathrm{GSF}}^{\omega_i,\omega_o}
=
\frac{1}{\omega_o}\,
\operatorname{diag}
\left(
\frac{1}{\omega_i} D_1 - L_1,\;
\frac{1}{\omega_i} D_2 - L_2,\;
\ldots,\;
\frac{1}{\omega_i} D_p - L_p
\right),
\end{equation}
where $\omega_i$ is the inner relaxation parameter for local updates in each processor, and $\omega_o$ is the outer relaxation parameter for global Jacobi-like updates at the processor boundaries \cite{Yang04}. Parallel versions of hybrid Gauss-Seidel backward (GSB) and Gauss-Seidel symmetric (GSS) can  similarly be constructed, and these hybrid smoothers are later denoted with their iteration operators $S^H_\text{GSF},\,S^H_\text{GSB},\,S^H_\text{GSS}$. A parallel weighted-Jacobi smoother can be constructed using a single relaxation parameter $\omega$ across the entire domain, and is denoted by $S_\text{Jacobi}$.
Instead of using relaxation parameters, an alternative way to improve convergence is by adding a diagonal matrix $D^{l_1} = \operatorname{diag}(d_{11},d_{22},\dots,d_{pp})$ to $B$, where $
d^{l_1}_{ii} = \sum_{j \in \Gamma} |a_{ij}|$ and $\Gamma$ are the sets of columns in the off-diagonal blocks of $A$. These are called \emph{$l_1$ smoothers},  compatible with the relaxation schemes mentioned earlier, and the respective iteration operators are denoted by $S^{l_1}_\text{GSF},\,S^{l_1}_\text{GSB},\,S^{l_1}_\text{GSS},\,S^{l_1}_\text{Jacobi}$. Each of these \emph{weighted smoothers} and \emph{$l_1$ smoothers} is applied lexicographically by default, but can also apply \emph{C-F smoothing}; that is, smoothing first on the coarse points and then on the fine points \cite{Baker2011}. 
\subsection{\ac{G3P}}
\label{sec:g3p}
Evolutionary algorithms (EAs) are a class of optimization techniques inspired by the principles of evolution by natural selection. \ac{GP} is a subclass of EAs focussed on evolving computer programs. The process begins with an initial population of $\rho_0$ randomly generated computer programs that incrementally become better over generations $t=0,\dots,t_{\max}$ by applying a fitness measure $f$. The population $\mathcal{P}^t$ at every generation $t$ has $\mu$ programs, with each program $x \in \mathcal{P}^t = \{x_1^t, \dots, x_\mu^t\}$ being evaluated on a fitness measure $f(x)$, over a set of problems $\mathcal{D}$. This is the most expensive step in the optimization, and hence the computation is distributed across MPI processes in a high-performance computing cluster. $N_{pop}$ is the number of MPI processes for the \emph{population-level parallelism}, wherein, individuals are distributed across multiple compute nodes, and $n_{p}$ is the number of MPI processes for the \emph{program-level parallelism}, wherein, each program is evaluated in parallel within a compute node. Using the population fitness collection $\mathcal{F}^t=\{\mathbf{f_1^t}, \dots, \mathbf{f_\mu^t\}}$, the fittest individuals are selected to produce offspring $\tilde{\mathcal{P}}^t$ of size $\lambda$ by applying a genetic operator $\mathcal{V}$, which either performs i) \emph{crossover} with probability $p_c$: exchanges parts of the program between two fit parents,  or ii) \emph{mutation} with probability $(1-p_c)$: introduces changes to an existing program for exploration. From the current population $\mathcal{P}^t$ and the offspring  $\tilde{\mathcal{P}}^t$ the next generation of $\mu$ programs is chosen with a suitable selection mechanism $\mathcal{S}$.  This process is repeated iteratively until the last generation $t_{max}$ \cite{GPfieldguide, GPIntro, Banzhaf2001, Orlov2009}. However, the initial population and genetic operations could generate invalid solutions within the population. When programs are required to conform to a specific structure or a predefined set of rules (for example, AMG programs) such information can be injected into the evolutionary process using CFGs \cite{Cremers1975}. Such a grammar $\mathcal{G}$ provides a formal system to embed a set of pre-defined rules that, when applied, generate \emph{valid} candidate programs. 
\ac{G3P} is an extension of traditional \ac{GP} systems that uses such a grammar $\mathcal{G}$ (\cref{alg:gggp}). $\mathcal{E}_{params}=\{ \mu, \lambda, \rho_0, \mathcal{S}, p_c, t_{\max}\}$ is the set of parameters used for the evolutionary algorithm. \ac{G3P} uses a population initialiser $\mathcal{I}$ that generates $\rho_0$ programs by applying rules from the grammar $\mathcal{G}$. Furthermore, the genetic operator $\mathcal{V}$ is augmented to $\mathcal{V_G}$, wherein crossover and mutation operations are restricted to respect the rules of the grammar and generate valid offspring. Using such a grammar $\mathcal{G}$ can help inject domain knowledge within GP systems. This improves search efficiency, helps discover better solutions, and guarantees the validity of generated programs \cite{Whigham1995GrammaticallybasedGP, Manrique2009}, in the sense of:  (i) \emph{syntactic correctness}, ensuring that the programs compile without errors, and (ii) adherence to  \emph{algorithmic structures} based on expert domain knowledge. In this paper, we formulate a grammar $\mathcal{G}$ for generating AMG methods in \emph{hypre} (\cref{fig:bnfboomeramg}). However, the numerical convergence of these methods is not guaranteed by the grammar, and instead, optimized by the GP search.


\begin{algorithm}[t]
\caption{Grammar-Guided Genetic Programming (G3P)}
\label{alg:gggp}
\begin{algorithmic}[1]

\STATE \textbf{Input:} $\mathcal{G,D,E}_{params},N_{pop},n_p$
\STATE \textbf{Output:} $\mathcal{P}^{t_{\max}}$

\STATE $\mathcal{P}^0 = \{x_1^0,\dots,x_{\rho_0}^0\} \gets \mathcal{I}(\mathcal{G},\rho_0) $ \hfill {\itshape // grammar-based initialization}
\STATE $\mathcal{F}^{0} = \{f_1^0,\dots,f_{\rho_0}^0\} \gets \textsc{Evaluate}(\mathcal{P}^{0},\mathcal{D},N_{pop},n_p)  $ \hfill {\itshape // measure population fitness} 

\FOR{$t=0$ to $t_{\max}-1$}
  \STATE $\tilde{\mathcal{P}}^t = \{x_1^t,\dots,x_{\lambda}^t\} \gets \mathcal{V}_{\mathcal{G}}(\mathcal{P}^t,\mathcal{F}^{t},p_c, \lambda) $ \hfill {\itshape // grammar-constrained offspring}
  \STATE $\tilde{\mathcal{F}}(t) \gets \textsc{Evaluate}(\tilde{\mathcal{P}}^t,\mathcal{D},N_{pop},n_p)$ \hfill {\itshape // measure offspring fitness}
  \STATE $P^{t+1} =\{x_1^{t+1},\dots,x_{\mu}^{t+1}\} \gets \mathcal{S}(\mathcal{P}^t,\tilde{\mathcal{P}}^t,\mathcal{F}^t,\tilde{\mathcal{F}}^t,\mu)$ \hfill {\itshape // select new population} 
  \STATE $\mathcal{F}^{t+1} = \textsc{Evaluate}(\mathcal{P}^{t+1})$ \hfill {\itshape // measure population fitness} 
\ENDFOR

\STATE \textbf{Return} $\mathcal{P}^{t_{\max}}$

\end{algorithmic}
\end{algorithm}

\color{black}

%% file: siamart_220329/contents/method.tex
\section{Method}
\label{sec:method}
Let $\mathcal{D} = \{(A_i,\mathbf{b_i})\}_{i=1}^N$ be a set of linear systems from a discretized parametric PDE and/or time-stepping sequence in a simulation code. Let $\mathcal{H}$ be a set of AMG algorithms generated from a grammar $\mathcal{G}$. $\mathbb{E}(h;A,\mathbf{b})$ is the \emph{wall-clock time} for solving a system $(A,\mathbf{b})$ using an algorithm $h$ on the target hardware. We want to find the fastest algorithm $h^\star \in \mathcal{H}$ for $\mathcal{D}$, which can be formally stated as

\begin{equation}
\label{eq:optprob}
h^\star
=
\arg\min_{h \in \mathcal{H}}
\sum_{(A, \mathbf{b})\in\mathcal{D}} \mathbb{E}(h;A,\mathbf{b}).
\end{equation}

We explore $\mathcal{H}$ using \ac{G3P} with a suitable fitness measure. A naive choice for the fitness would be the \emph{wall-clock time} $\mathbb{E}$, but we define a vector-valued fitness $\mathbf{F} = [T/N, \rho]^T$
and perform a multi-objective optimization. Considering the \emph{cost per iteration} $T/N$ and \emph{convergence rate} $\rho$ as two separate objectives, creates diversity in the population, encourages exploration, and generates better
solutions \cite{MO1,MO2}. 
G3P can be viewed as the following operation
\begin{equation}
    \textsc{G3P} : (\tilde{\mathcal{D}}, \mathcal{G}, \mathbf{F}, \mathcal{E}_{params})
\;\longrightarrow\;
\mathcal{P}^\star \subset \mathcal{P}^{t_{\max}},
\end{equation}
where, $\tilde{\mathcal{D}}$ is a set of cheap proxy problem(s) for $\mathcal{D}$, with $|\tilde{\mathcal{D}}|\ll |\mathcal{D|}$, hence directly impacting the optimization cost (lines 4 and 7 in \cref{alg:gggp}). From the Pareto front or a set of optimal solutions $P^\star\subset \mathcal{P}^{t_{\max}}$, several algorithms are hand-picked by the user to find the best algorithm $\tilde{h}$ for solving $\mathcal{D}$ such that

\begin{equation}
    \tilde{h}
\;=\;
\arg\min_{h \in \mathcal{P}^\star}
\sum_{(A,\mathbf{b})\in\mathcal{D}}
|
\mathbb{E}(h;A,\mathbf{b})
-
\mathbb{E}(h^\star;A,\mathbf{b})|
\;=\;
\arg\min_{h \in \mathcal{P}^\star}
\sum_{(A,\mathbf{b})\in\mathcal{D}}
\mathbb{E}(h;A,\mathbf{b}).
\end{equation}

\subsection{Flexible \ac{AMG} cycles}
\label{sec:flexibleAMG}
We extend equation \cref{eq:MGoperator} and define a \emph{flexible} multigrid operator as

\begin{equation}
    M_l^\text{flex} = \prod_j T_l^j, \quad T_l^j \in \{S_l^j,C_l^j\}, \quad l>0,
\end{equation}
representing an 
arbitrary sequence of smoothing and coarse-grid corrections steps. Here, $S_l^j$ is any valid smoother, $C_l^j=I - \alpha^{(j)}P_l(B_{l-1}^j)^{-1}R_lA_l$ is the coarse-grid correction step with a scaling factor $\alpha$, and $(B_{l-1}^j)^{-1}$ is the application of a flexible multigrid $M_{l-1}^{\text{flex},j}=I-(B_{l-1}^j)^{-1}A_{l-1}$ to approximate the coarse system $(A_{l-1})^{-1}$. This can be extended to multiple levels, to obtain a multi-level \emph{flexible} multigrid operator.

\textit{Why is genetic programming required?} Let us restrict ourselves to a V-cycle-like topology with \emph{flexible smoothing}. Let $\mathcal{S}$ be a finite set of admissible smoothers, $m = |\mathcal{S|}$. Let $\nu_{max}$ be the maximum number of pre-/ post-smoothing steps allowed in $M_l^{flex}$, that is, $\nu_l^{pre},\nu_l^{post} \leq \nu_{max}$. The number of distinct smoothing sequences at each level is $N_l^{smooth} = (\sum_{\nu = 1}^{\nu_{max}}m^\nu)^{2} = (\frac{m^{\nu_{max}+1}-1}{m-1})^2 \approx  m^{2\cdot\nu_{max}}$. Thus, the number of AMG cycles $N_{cycles}$ for a hierarchy with $L+1$ levels is
\begin{equation}
    \label{eq:searchspace}
    N_{cycles} \approx m^{2 \cdot L \cdot \nu_{max}}.
\end{equation}
For example, with, $L=5, \nu_{max} =3, m=6$, $N_{cycles} \approx 2 \times 10^{23}$. This is a highly conservative estimate assuming a fixed cycle topology, and is already intractable to design manually by hand. Hence, we use genetic programming to automatically design efficient, flexible AMG cycles.

\subsection{Design constraints}
\label{sec:designconstraints}
Since the search space scales exponentially with the number of grid levels $L$ (equation \cref{eq:searchspace}), we limit \emph{flexible cycling} only for the top $N_\mathrm{flex}$ levels $L, L-1, \dots ,l_\text{std}$ where $l_\mathrm{std}=L-N_\text{flex} +1$. A standard V-cycle operator $M^V$ is used for the remaining levels. This can be formally defined as 

\begin{equation}
    \label{eq:FlexMGmodified}
    M_l^\text{flex} = 
    \begin{cases}
    \prod_j T_l^j, \quad T_l^j \in \{S_l^j,C_l^j\}, \quad l>l_{std}, \\
    M_l^V, \quad 0<l \leq l_{std}.
    \end{cases}
\end{equation}

In addition to reducing the optimization cost, the modified formulation \cref{eq:FlexMGmodified} also allows us to scale our generated solvers to changes in the number of grid levels, for example, during weak scaling (\cref{subsubsec:evalweakscaling})  or solving different linear systems in a time-stepping code (\cref{subsec:aresweakscaling,subsec:arestimestepping}). We choose $N_\mathrm{flex}=5$, a set of smoothers 
\begin{equation}
   \mathcal{S}=\underbrace{\{S_{\mathrm{GSF}}^{H},S_\mathrm{GSB}^{H},S_\mathrm{Jacobi}\}}_\text{weighted smoothers} \cup \underbrace{\{S_{\mathrm{GSF}}^{l_1},S_\mathrm{GSB}^{l_1},S_\mathrm{Jacobi}^{l_1}\}}_\text{$l_1$ smoothers},
   \label{eq:smootherset}
\end{equation}

that can be applied in either \emph{lexicographical} or \emph{C-F} ordering, and relaxation parameters for weighted smoothers $\omega_i,\omega_o,\omega$ and scaling factors for coarse-grid correction $\alpha$ sampled from $\{0.1,0.15, \dots,1.9\}$. The V-cycle from $l=l_\mathrm{std}$ uses $S_\mathrm{GSF}^{l_1}$ for pre-smoothing, $S_\mathrm{GSB}^{l_1}$ for post-smoothing, and Gaussian elimination at $l=0$. The AMG hierarchy is generated using HMIS coarsening strategy \cite{DeSterck2006}, an extended long-range interpolation operator \cite{DeSterck2007}, and a strength threshold of $0.25$.

\subsection{Context-free grammar for \emph{flexible} AMG}
We formulate a \ac{CFG} comprising a list of production rules (\cref{fig:bnfboomeramg}). Each production rule contains a production symbol $\bnfpn{ . }$ (to the left) with the corresponding expression (to the right). The production rules are applied from the start symbol, $\bnfpn{$G$}$, recursively until the resultant expression contains only \emph{terminals}. The final expression maps to a unique instance of an \ac{AMG} cycle.
$\bnfpn{$s_L$}$, $\bnfpn{$s_{l_\mathrm{std}}$}$, and $\bnfpn{$s_l$}$ represent a \emph{state of approximation} in the \ac{AMG} cycle at the finest level, the last flexible level and intermediate levels $l_\mathrm{std}<l<L$, respectively (Fig. \ref{fig:bnfboomeramg}, left).
For example, in order to reach the state, $\bnfpn{$s_L$}$, one of the following operations must have been applied:  (i) any sort of smoothing at $l=L$,  (ii) coarse-grid correction from $l=L-1$, or (iii) starting initial guess (cycle terminates). Similarly, for the intermediate state levels, $\bnfpn{$s_l$}$, apart from smoothing or coarse-grid correction, \emph{restriction} from level $l+1$ is allowed, and at $\bnfpn{$s_{l_\mathrm{std}}$}$  a standard recursive multigrid is applied. These production rules interweave different grid levels in the \ac{AMG} hierarchy with valid multigrid operations, generating \emph{flexible} cycle structures. The smoothing choices, smoothing ordering, relaxation weights, and coarse-grid correction scaling factors are specified as \emph{terminal symbols},  (\cref{fig:bnfboomeramg}, right), that is, concrete algorithmic choices which cannot be expanded further in the grammar. 
\hspace{15cm}
\begin{minipage}[!htbp]{0.64\textwidth}
\flushright
\setlength{\fboxsep}{1pt} 
\setlength{\fboxrule}{0.7pt} 
\framebox{%
  \begin{minipage}{0.93\textwidth}
  \footnotesize
    \begin{bnf*}
    \bnfprod{$G$}{\bnfpn{$s_L$}} \\
    \bnfprod{$s_L$}{\textsc{$l_1$Relax}(\bnfpn{$S^{l_1}$},\bnfpn{$\mathcal{R}$},\bnfpn{$s_L$})\\
    \bnfmore}{\textsc{HybridRelax}(\bnfpn{$S^H$},\bnfpn{$w_i$},\bnfpn{$w_o$},\bnfpn{$\mathcal{R}$},\bnfpn{$s_L$})}\\
    \bnfmore{\textsc{JacobiRelax}(S_\mathrm{Jacobi},\bnfpn{$w$},\bnfpn{$\mathcal{R}$},\bnfpn{$s_L$})}\\
    \bnfmore{\textsc{CoarseGridCorrection}(\bnfpn{$\alpha$},\bnfpn{$s_{L-1}$})}\\
    \bnfmore{\bnftd{\textsc{InitialGuess}}}\\
    \bnfprod{$s_{l_\mathrm{std}}$}{\textsc{RestrictAndSolve}(M^V,\bnfpn{$s_{l_\mathrm{std}+1}$})}\\
    \bnfprod{$s_l$}{\textsc{$l_1$Relax}(\bnfpn{$S^{l_1}$},\bnfpn{$\mathcal{R}$},\bnfpn{$s_l$})}\\
    \bnfmore{\textsc{HybridRelax}(\bnfpn{$S^H$},\bnfpn{$w_i$},\bnfpn{$w_o$},\bnfpn{$\mathcal{R}$},\bnfpn{$s_l$})}\\
    \bnfmore{\textsc{JacobiRelax}(S_\mathrm{Jacobi},\bnfpn{$w$},\bnfpn{$\mathcal{R}$},\bnfpn{$s_l$})}\\
    \bnfmore{\textsc{Restrict}(\bnfpn{$s_{l+1}$})}\\
    \bnfmore{\textsc{CoarseGridCorrection}(\bnfpn{$\alpha$},\bnfpn{$s_{l-1}$})}\\
    \end{bnf*}
  \end{minipage}
  }
\end{minipage}%
\hfill
\begin{minipage}[!htbp]{0.36\textwidth}
\flushleft
\setlength{\fboxsep}{5pt} 
\setlength{\fboxrule}{0.7pt} 
\framebox{%
  \begin{minipage}{0.85\textwidth}
  \underline{\emph{Terminals}}
  \footnotesize
    \begin{bnf*}
    \bnfprod{$S^{l_1}$}{\bnftd{$S^{l_1}_{GSF},S^{l_1}_{GSB}, S^{l_1}_\mathrm{Jacobi}$}} \\
    \bnfprod{$S^{H}$}{\bnftd{$S^{H}_{GSF},S^{H}_{GSB}$}} \\ \\
    \bnfprod{$\mathcal{R}$}{\bnftd{Lexicographic}} \\
    \bnfmore{\bnftd{\text{C-F} smoothing}}\\ \\
    \bnfprod{$w_i$}{\bnftd{$0$ \(|\)  $0.15$ \(|\) ... \(|\)  $1.9$}}\\
    \bnfprod{$w_o$}{\bnftd{$0$ \(|\)  $0.15$ \(|\) ... \(|\)  $1.9$}}\\
    \bnfprod{$w$}{\bnftd{$0$ \(|\)  $0.15$ \(|\) ... \(|\)  $1.9$}}\\
    \bnfprod{$\alpha$}{\bnftd{$0$ \(|\)  $0.15$ \(|\) ... \(|\)  $1.9$}}\\
    \end{bnf*}
    \vspace{-7pt}
  \end{minipage}
}
\end{minipage}
\captionof{figure}{Production rules for the generation of \emph{flexible} BoomerAMG methods.}
\label{fig:bnfboomeramg}

\subsection{Software}
\begin{figure}[!htbp]
  \centering
  \includegraphics[width=\linewidth]{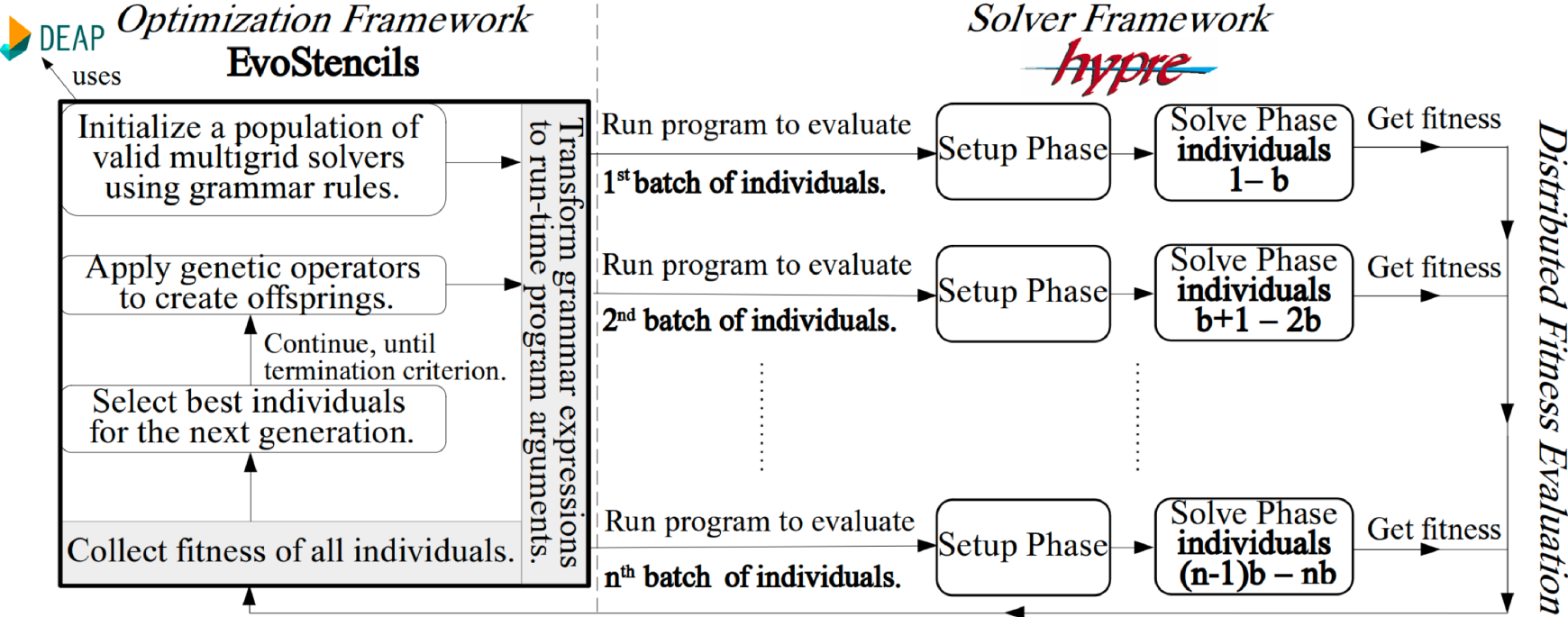}
  \caption{An overview of the software setup for the automated \ac{AMG} design.}
  \label{fig:optpipeline}
\end{figure}

The \ac{CFG} has been implemented using \emph{EvoStencils}\footnote{https://github.com/jonas-schmitt/evostencils}, a Python package for the automated design of multigrid methods. Additional interfaces have been added and implemented in \emph{hypre} to incorporate \emph{flexible cycling} in BoomerAMG, the AMG implementation in \emph{hypre}\footnote{https://github.com/dinesh-parthasarathy/hypre}. The \ac{CFG} generates \ac{AMG} expressions, which are then transformed into corresponding BoomerAMG inputs $\mathbf{\theta}$ for these interfaces. The linear systems are imported via the \emph{Linear-Algebraic System Interface} in \emph{hypre}\footnote{https://hypre.readthedocs.io/en/latest/ch-ij.html}  and solved with BoomerAMG arguments $\mathbf{\theta}$ (\cref{alg:evaluate}). Evolutionary algorithms from the DEAP library \cite{DEAP_JMLR2012} are used, and the \emph{population-level parallelism} $N_{pop}$ is implemented using Python bindings for MPI\footnote{https://mpi4py.readthedocs.io/en/stable/} (\cref{fig:optpipeline}). To eliminate redundant AMG setup times, a batch of individuals is lumped within a single program with a common AMG setup. The evolutionary parameters $\mathcal{E}_{params}$ are set to $\mu,\lambda=256,\rho_{0}=2048$, the NSGA-II algorithm \cite{NSGA} for selection $\mathcal{S}, p_c=0.7, t_\mathrm{max}=100$.

 \begin{algorithm}[t]
\caption{\textsc{Evaluate}$(\mathcal{P},\tilde{\mathcal{D}}, N_{pop}, n_p)$}
\label{alg:evaluate}
\begin{algorithmic}[1]

\STATE  $\mathcal{P} = \bigcup_{k=1}^{N_{pop}} \mathcal{P}_k$ \hfill {\itshape // split population $\mathcal{P}$ across $N_{pop}$ processes}
\STATE $\mathcal{F} = \emptyset$ \hfill {\itshape // fitness collection for the population}

\FORALL{$k = 1,\dots,N_{pop}$ \textbf{in parallel}}
  \FORALL{$x \in \mathcal{P}_k$}
    \STATE $\boldsymbol{\theta} \gets \textsc{GetHypreArgs}(x)$ \hfill {\itshape // convert genotype to phenotype}
\STATE $(T, N, \rho)
\gets
\frac{1}{|\tilde{\mathcal{D}}|}
\sum_{(A,b)\in\tilde{\mathcal{D}}}
\textsc{RunHypre}(\boldsymbol{\theta},A,b, n_p)$
    \STATE $\mathbf{f_x} = \left[\frac{T}{N}, \rho \right]^T$
    \STATE $\mathcal{F} \gets \mathcal{F} \cup \{\mathbf{f_x}\}$
  \ENDFOR
\ENDFOR

\STATE \textbf{Return} $\mathcal{F}$

\end{algorithmic}
\end{algorithm}

%% file: siamart_220329/contents/experiments.tex
\section{Automated AMG solver design for a 3D anisotropic Poisson problem}
\label{sec:amgsolver}
First, we use our methodology to design AMG solvers for the 
anisotropic Poisson problem on a unit cube domain:
\begin{equation}
\begin{aligned}
- \nabla \cdot ( D \nabla u ) &= f 
\quad \text{in } \Omega = (0,1)^3, \\
\text{with    }u &= 0 \ \ \text{on} \ \partial\Omega ,\\
\text{where    }D &= \mathrm{diag}(c_1, c_2, c_3),
\quad c_1, c_2, c_3 > 0.
\end{aligned}
\label{eq:anisotropicpoisson}
\end{equation}
The problem is discretized on an equally spaced 3D mesh having $N_d^3$ unknowns. For each PDE formulation with $\phi=(c_1,c_2,c_3,N_d)$, the matrix $A(\phi) \in \mathbb{R}^{N_d^3 \times N_d^3 }$ is obtained using a finite-difference scheme with a standard 7-point Laplacian stencil, leading to a linear system of the form (\ref{eq:linear}).
The resulting matrix is sparse, symmetric positive definite (SPD), and has a banded structure with at most 7 nonzero entries per row. From \cref{eq:anisotropicpoisson}, we are interested in the problem set
\begin{equation}
\mathcal D
=
\left\{
A(\phi)
\;\middle|\;
\phi = (c_1,c_2,c_3,N_d),\;
(c_1,c_2,c_3)\in[10^{-5},1]^3,\;
N_d\in\{100, 101 \dots,1000\}
\right\},
\end{equation}
for automated AMG design.
We set $f=0$, use a random initial guess, and run the AMG solver for $N$ iterations until $\lVert r^{(N)}\rVert_2 \leq \epsilon=10^{-8}$. AMG is designed with \ac{G3P} for $\mathcal{D}$ using a proxy $\tilde{D}=\{A(\phi_{proxy}):\phi_{proxy}=(0.001,1,1,100)\}$ solved over $n_p=8$ MPI processes (See line 6, \cref{alg:evaluate}). 

\subsection{Reference solvers}
\label{subsec:refamgsolvers}
To benchmark the optimization results, we use a set of reference \ac{AMG} solvers tuned for $\tilde{\mathcal{D}}$. Only V-cycles are considered, as W-cycles with the default parameters were found to be not as competitive. For a V-cycle, all possible choices of pre-/post-smoothers, that is,  $S^{pre} \in \mathcal{S} \,\cup\{S_\text{GSS}^{H},\,S_\text{GSS}^{l_1}\},\, S^{post} \in \mathcal{S}\,\cup\{S_\text{GSS}^{H},\,S_\text{GSS}^{l_1}\}$ (refer \cref{eq:smootherset}), were evaluated.
Among \emph{weighted smoothers}, the relaxation weights, $\omega_i, \omega_o,\omega$, in symmetric smoothers were obtained by applying Lanczos or CG iterations as presented in \cite{Yang04}, while an exhaustive search was performed for the non-symmetric variants. They were further fine-tuned for different relaxation ordering; C-F and lexicographic relaxation, and for different pre-/post-smoothing steps, $\nu_1,\,\nu_2$. The resulting six fastest \ac{AMG} solvers alongside the default BoomerAMG configuration, are selected as reference solvers (Table \ref{tab:refsolvers}).

\begin{table}[!htbp]
    \centering
\begin{tabular}{|l||l|l|l|l|l|l|l|}
\toprule
 solver &    ordering & $S^{pre}$ & $\nu_1$ & $S^{post}$ & $\nu_2$ & $w_i$ & $w_o$\\
\midrule
\textbf{default} & lexicographic &  $S^{l_1}_\text{GSF}$ &         1 &   $S^{l_1}_\text{GSB}$ &          1 &   n/a &   n/a  \\
 \textbf{tuned 1} &  C-F &  $S^{l_1}_\text{GSF}$ &         1 &   $S^{l_1}_\text{GSF}$ &          1 &   n/a &   n/a  \\
\textbf{tuned 2} &  C-F &    $S^{H}_\text{GSF}$ &         1 &      $S^{H}_\text{GSF}$ &          1 &   1.1 &   0.9  \\
\textbf{tuned 3} & lexicographic &  $S^{l_1}_\text{GSF}$ &         1 &   $S^{l_1}_\text{GSF}$ &          1 &   n/a &   n/a \\
\textbf{tuned 4} &  C-F &    $S^{l_1}_\text{Jacobi}$ &         1 &     $S^{l_1}_\text{Jacobi}$  &          1 &   n/a &   n/a  \\
\textbf{tuned 5} & lexicographic &     $S^{H}_\text{GSF}$ &         1 &      $S^{H}_\text{GSF}$ &          1 &   1.1 &   0.9  \\
\textbf{tuned 6} & lexicographic &  $S^{l_1}_\text{GSF}$ &         2 &   $S^{l_1}_\text{GSB}$ &          1 &   n/a &   n/a  \\
\bottomrule
\end{tabular}
\caption{BoomerAMG parameters for default and tuned reference \ac{AMG} solvers.}
    \label{tab:refsolvers}
\end{table}
\subsection{Performance Evaluation and Generalizability}

\label{sec:eval}
 \begin{figure}[!htbp]
    \centering
    \begin{tabular}{|@{}c@{}|@{}c@{}|}
    \toprule
      \includegraphics[width=0.49\textwidth]{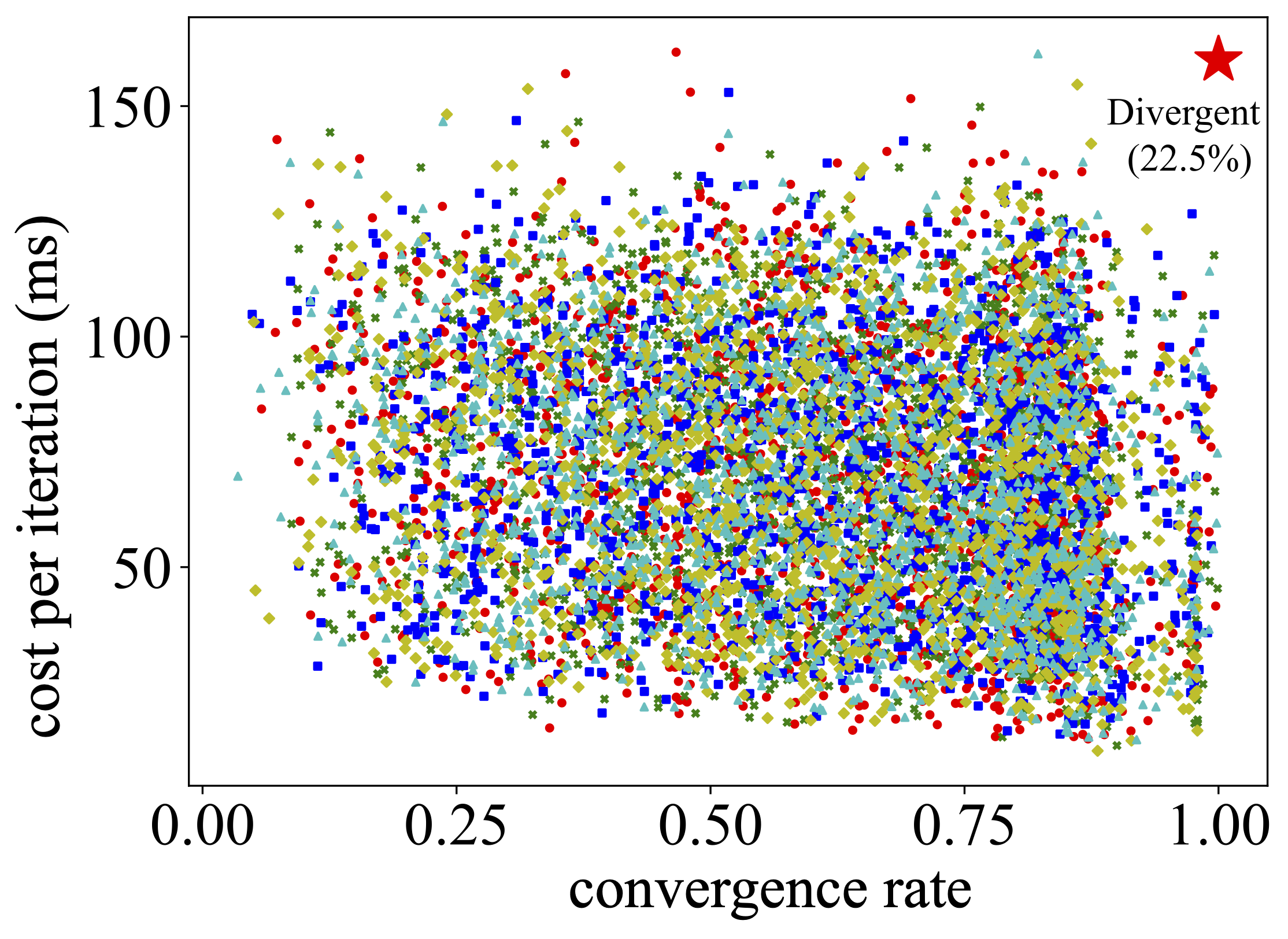} &
      \includegraphics[width=0.49\textwidth]{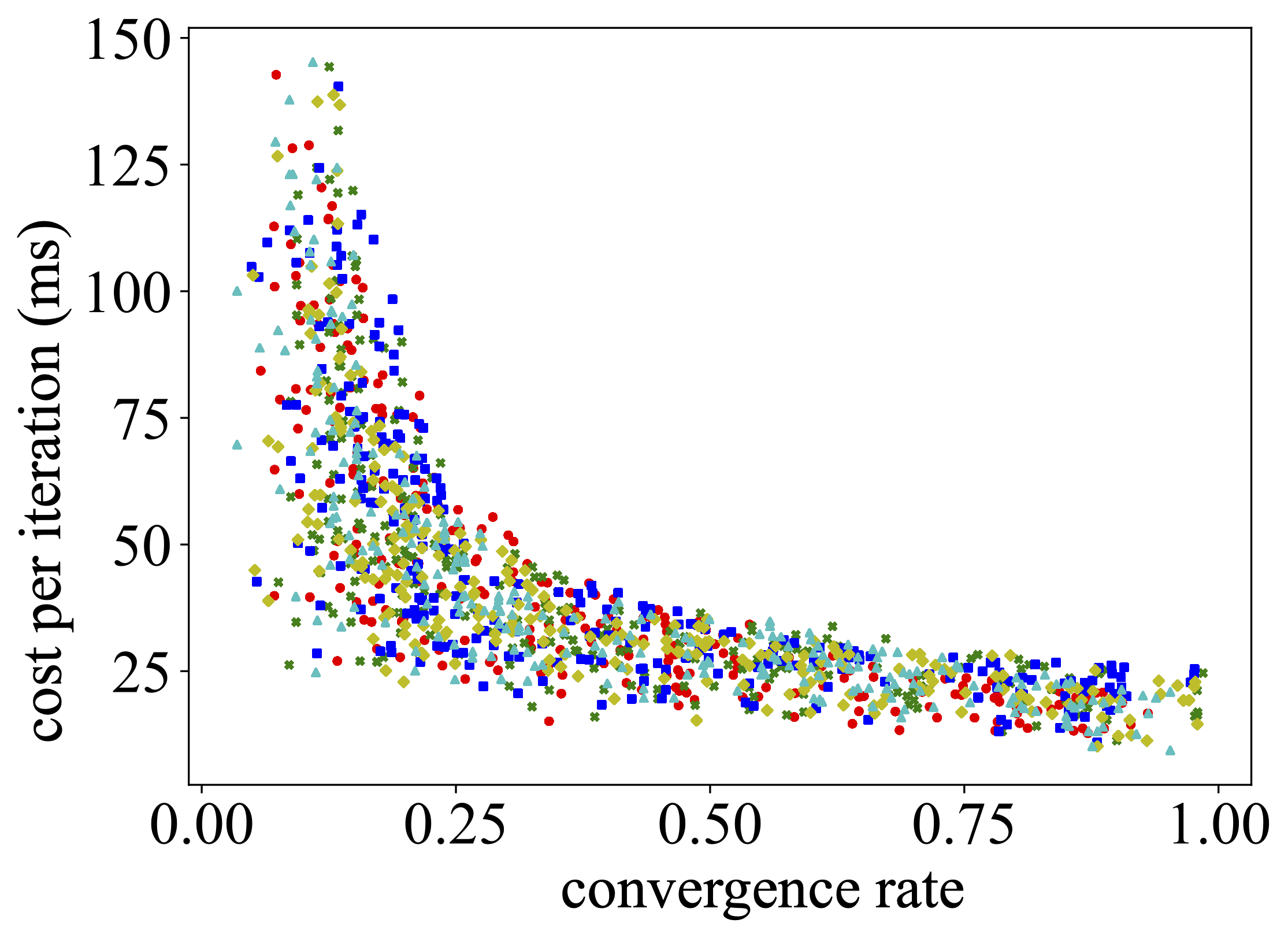} \\
      \midrule
      \includegraphics[width=0.49\textwidth]{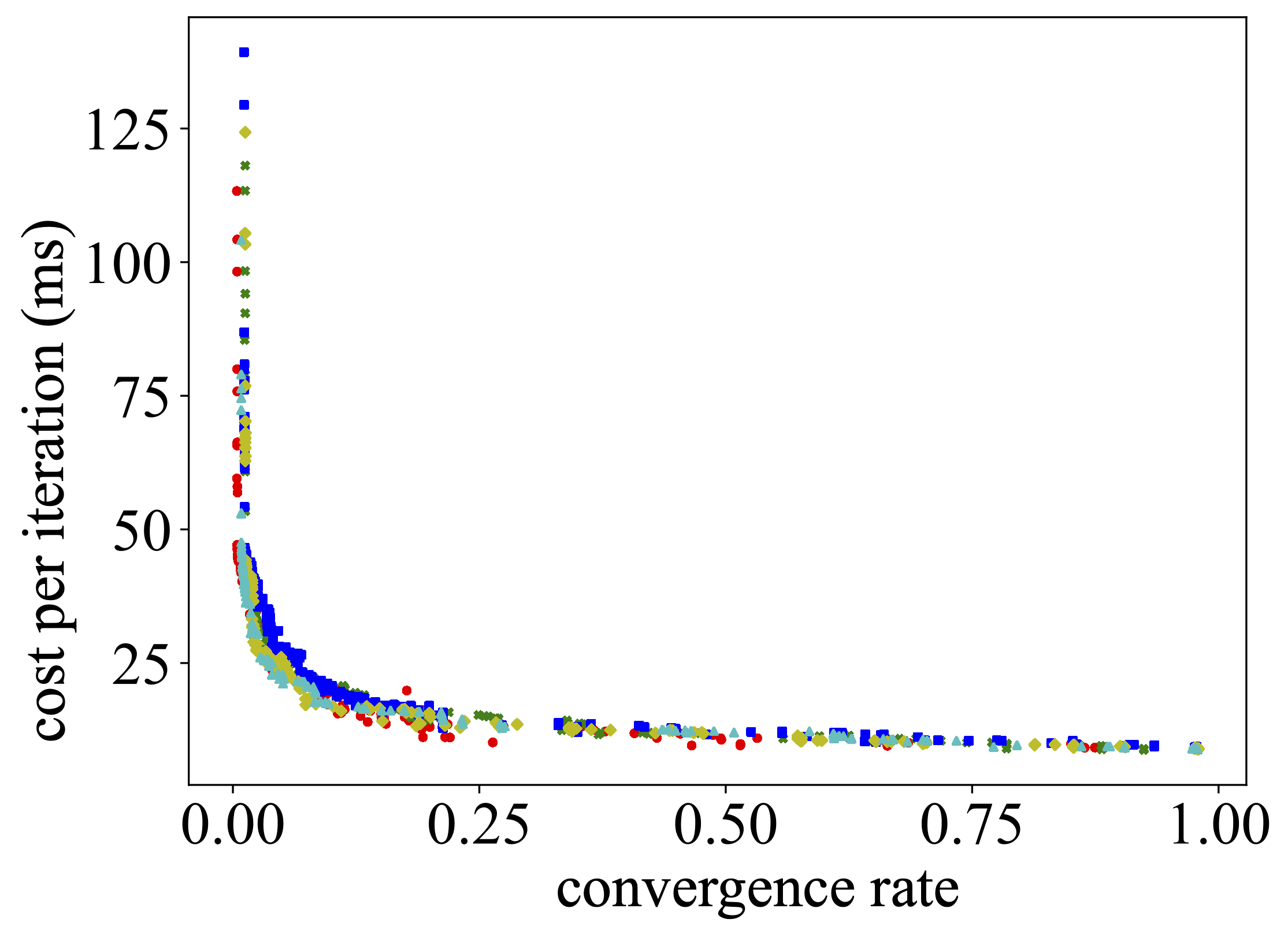} &
      \includegraphics[width=0.49\textwidth]{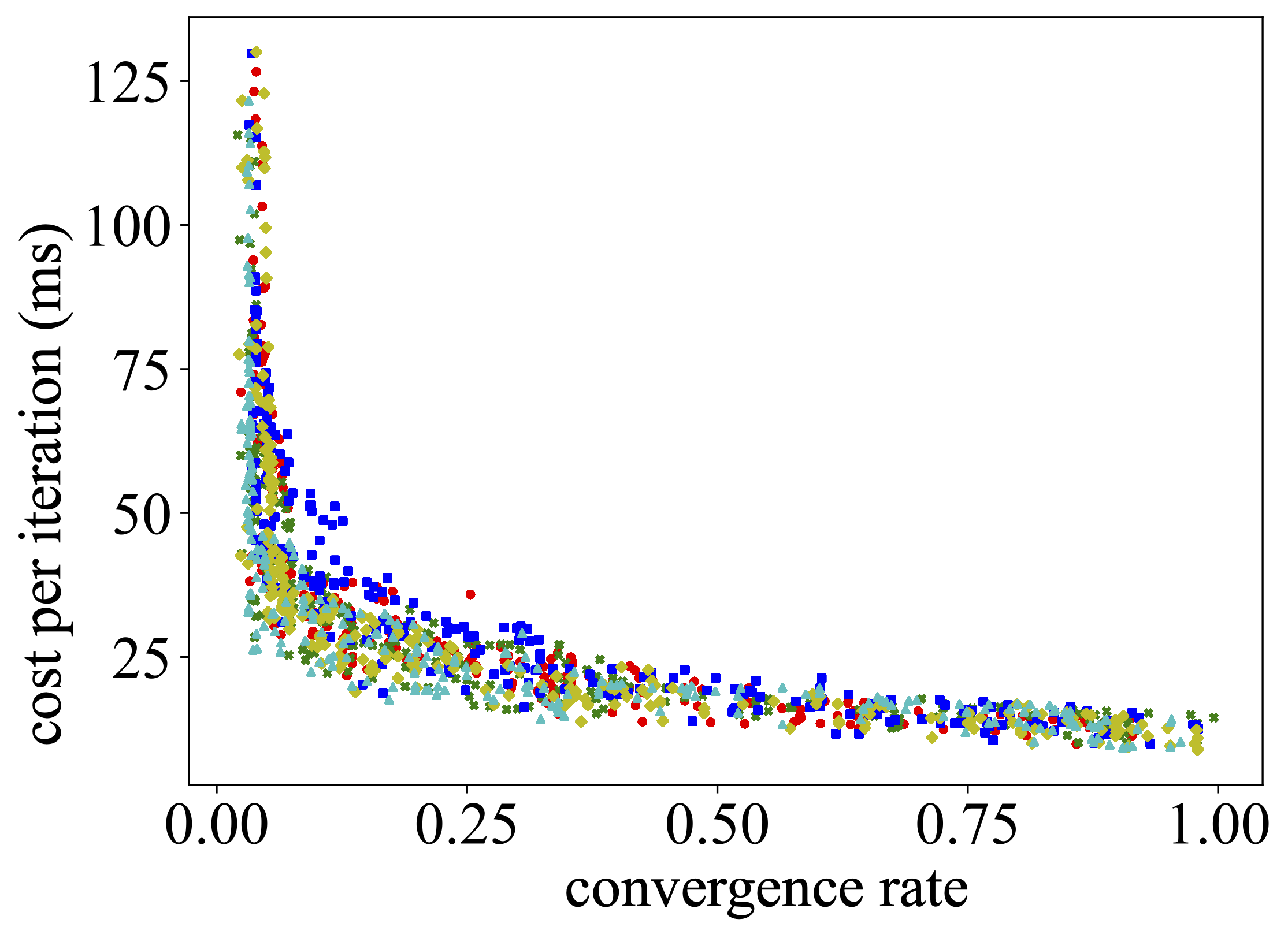} \\
          \bottomrule
    \end{tabular}
    \caption{Evolution of AMG individuals with respect to \textit{cost per iteration} and \textit{convergence rate}, progressing from an initial random population to the final generation (clockwise  
    from top left: \textit{initial population}, \textit{generation 1}, \textit{generation 10}, \textit{generation 100}). The dots represent the individuals, and are colored separately, representing each of the five independent runs of the G3P algorithm.}
    \label{fig:evolution}
  \end{figure}
\begin{figure}[!htbp]
    \centering
    \begin{subfigure}[t]{0.72\textwidth}
        \centering
        \includegraphics[width=\linewidth]{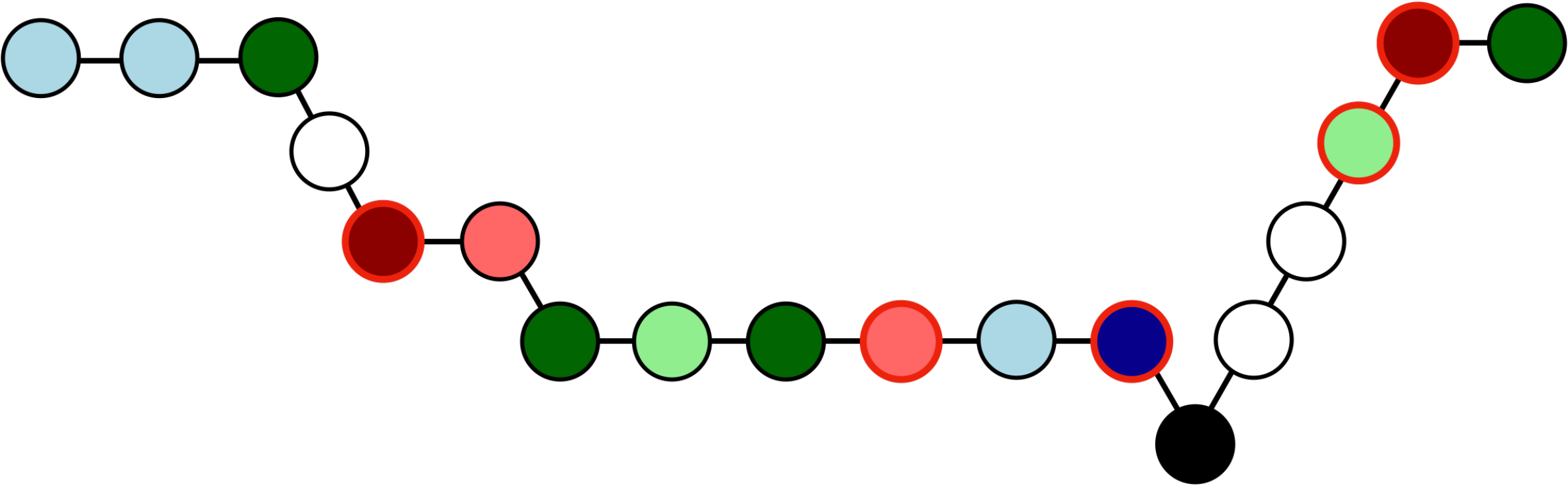}
        \caption{GP-22}
        \label{fig:GP22}
    \end{subfigure}
    \hfill
    \begin{subfigure}[t]{0.25\textwidth}
        \centering
        \includegraphics[width=\linewidth]{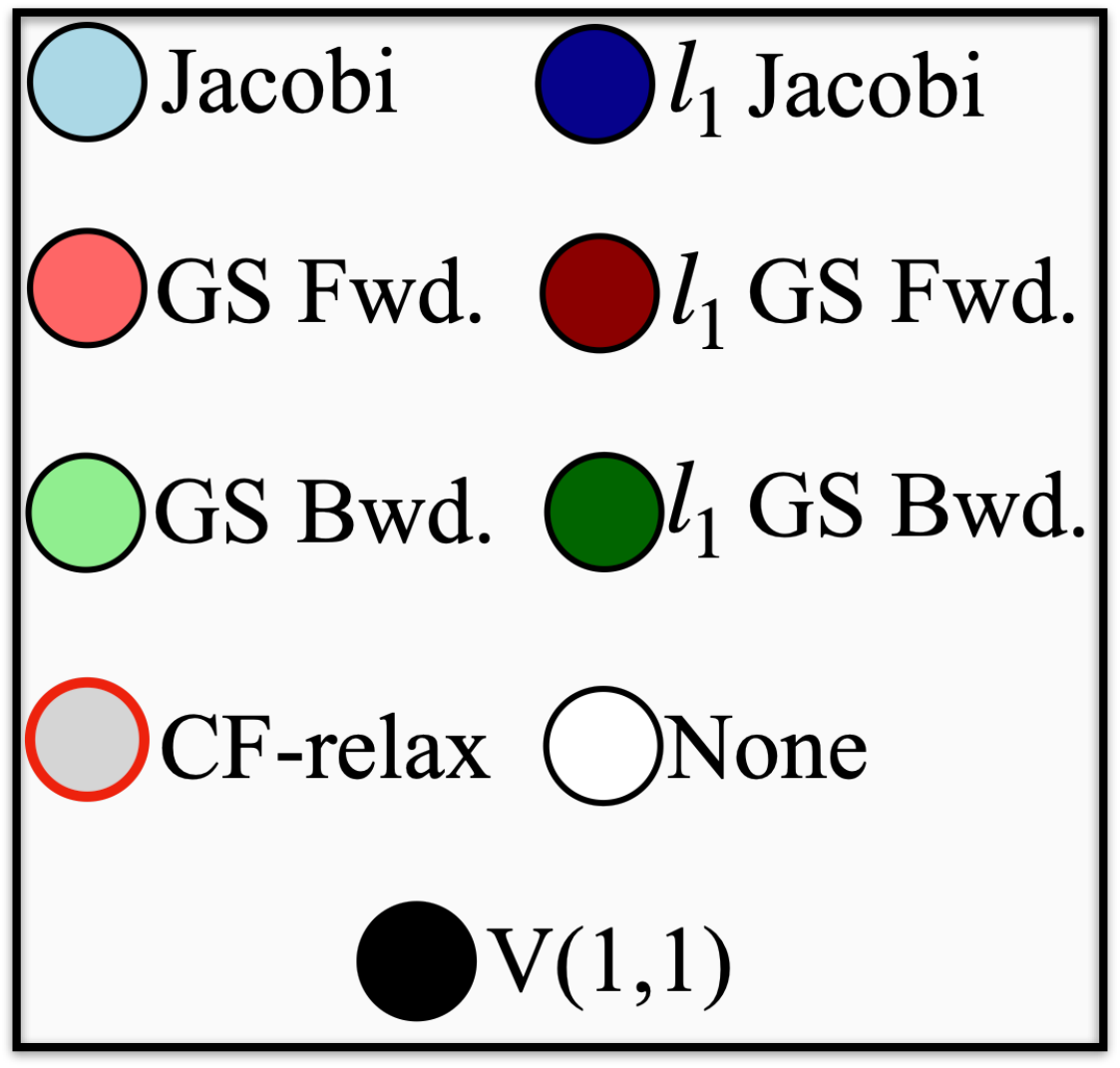}
    \end{subfigure}
    \begin{subfigure}[t]{0.38\textwidth}
        \centering
        \includegraphics[width=\linewidth]{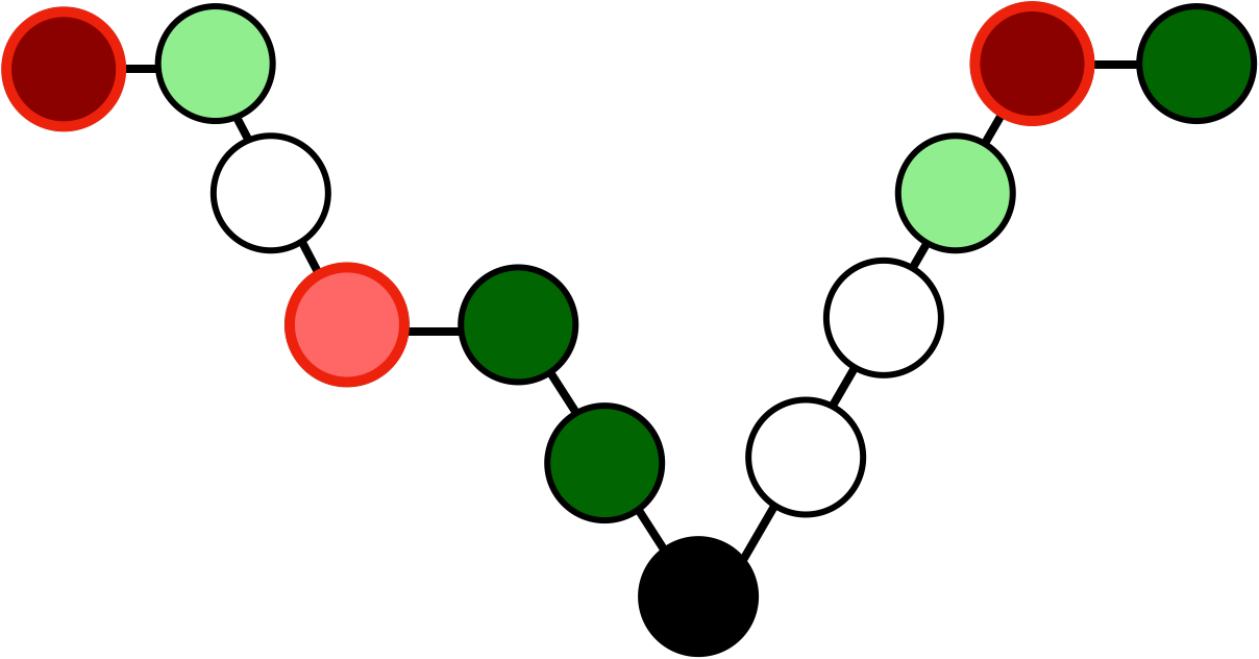}
        \caption{GP-27}
        \label{fig:GP27}
    \end{subfigure}
    \hfill
    \begin{subfigure}[t]{0.45\textwidth}
        \centering
        \includegraphics[width=\linewidth]{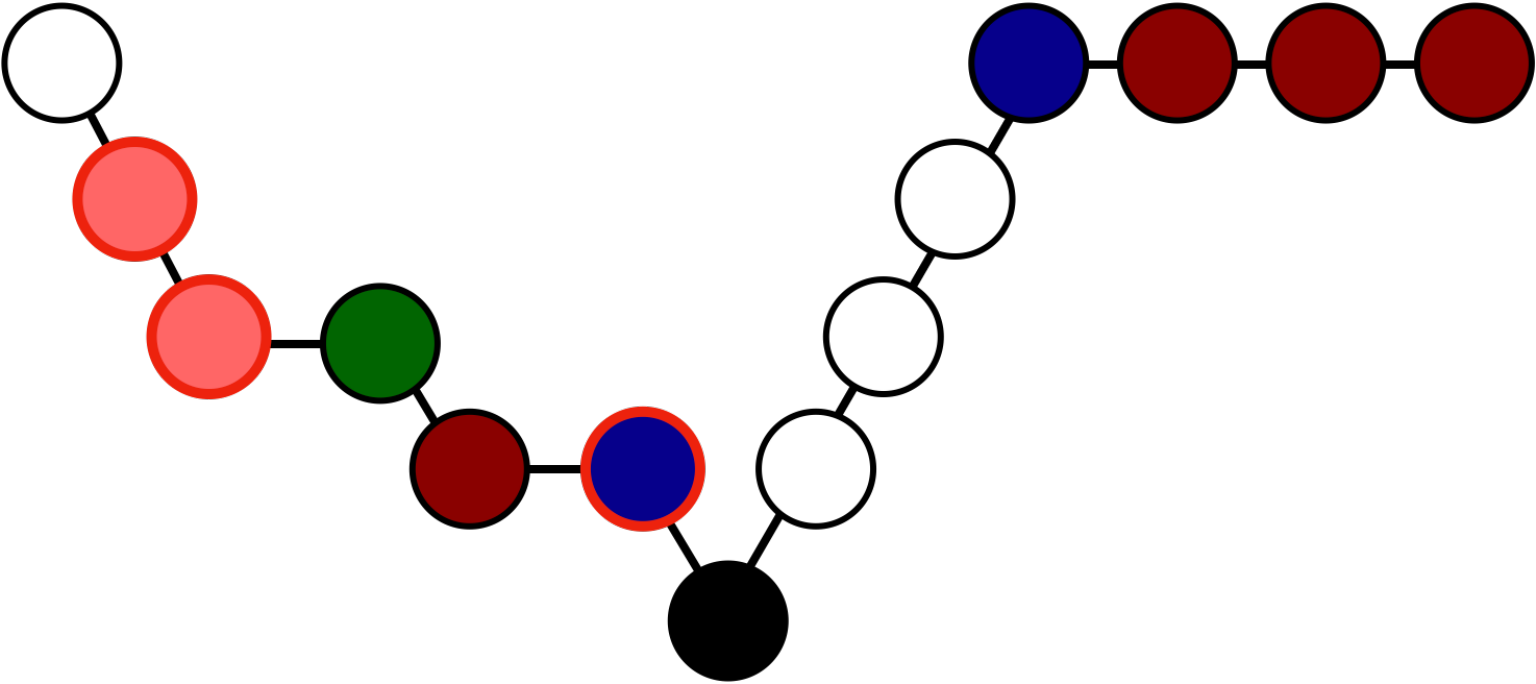}
        \caption{GP-28}
        \label{fig:GP28}
    \end{subfigure}
    \begin{subfigure}[t]{0.3\textwidth}
        \centering
        \includegraphics[width=\linewidth]{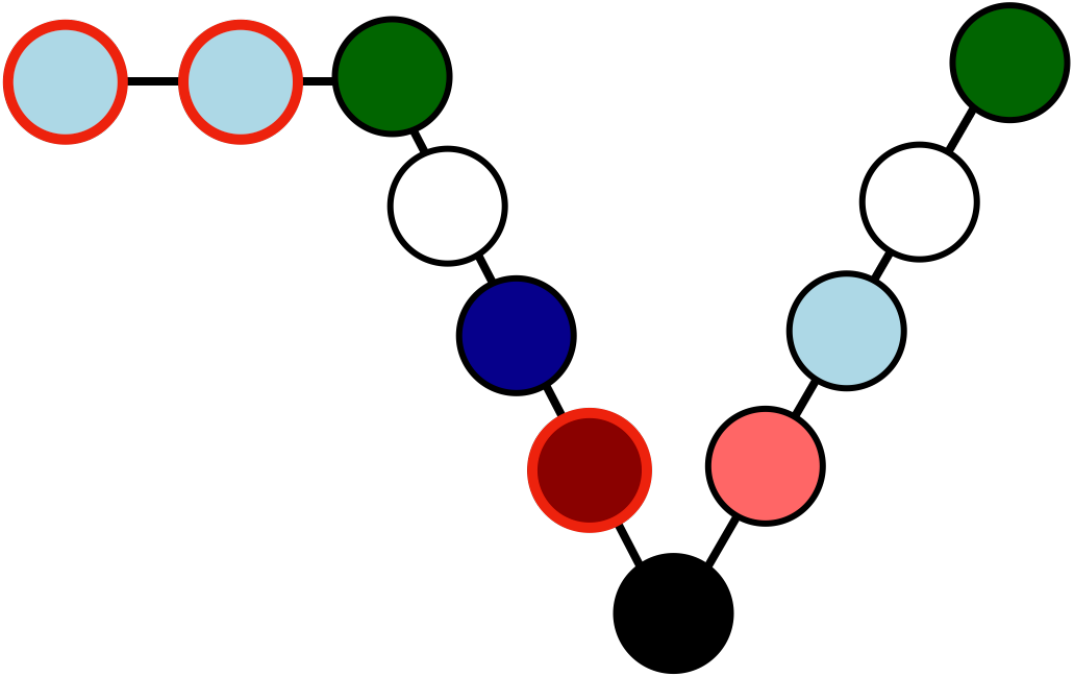}
        \caption{GP-31}
        \label{fig:GP31}
    \end{subfigure}
    \hfill
    \begin{subfigure}[t]{0.3\textwidth}
        \centering
        \includegraphics[width=\linewidth]{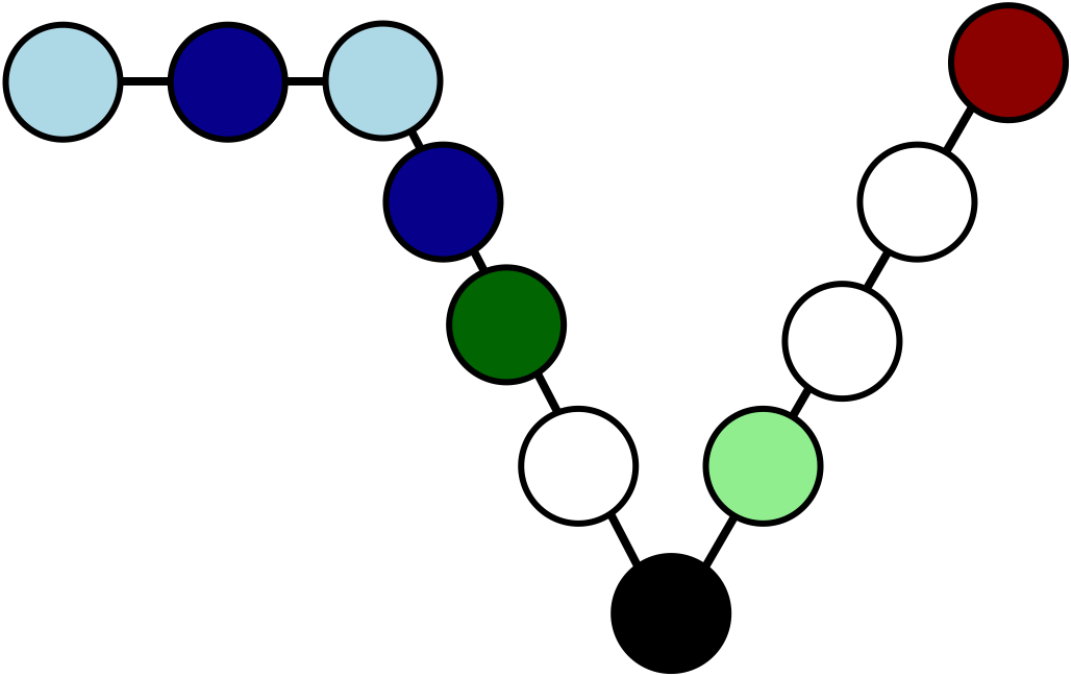}
        \caption{GP-33}
        \label{fig:GP33}
    \end{subfigure}
    \hfill
    \begin{subfigure}[t]{0.26\textwidth}
        \centering
        \includegraphics[width=\linewidth]{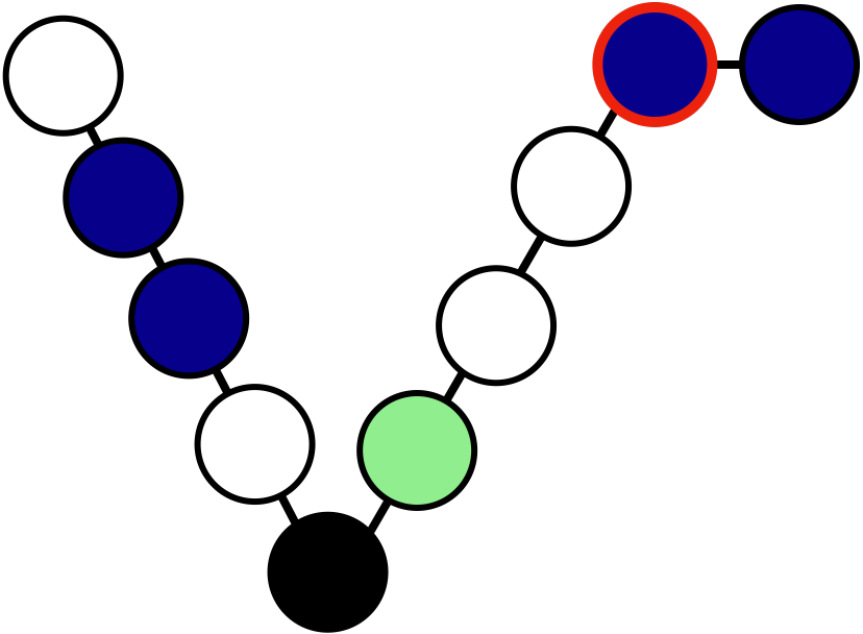}
        \caption{GP-36}
        \label{fig:GP36}
    \end{subfigure}

    \caption{AMG cycling structure for the selected GP solvers, excluding the depiction of weighting parameters $\omega_i$, $\omega_0$, $\omega$, and $\alpha$.}
    \label{fig:GPsolvers}
\end{figure}

  The optimization starts with a random collection of AMG solvers generated from the grammar rules, and many of them diverge or converge slowly (top-left, Fig. \ref{fig:evolution}). Applying genetic operators (crossover, mutation) and downselecting for the most fit individuals to this initial population produces a collection of cheaper AMG solvers with faster convergence (top-right, Fig. \ref{fig:evolution}). Repeating this iteratively evolves a population of solvers aligned/clustered close to the Pareto front. This optimization is independently repeated five times, each corresponding to different point colorings, demonstrating reproducibility (Fig. \ref{fig:evolution}). Six \ac{GP} solvers are picked from the final population (\cref{fig:GPsolvers}) and 
evaluated for generalization to the problem set $\mathcal{D}$ by picking problems with different anisotropies (\cref{subsubsec:evalanisotropy}) and different number of unknowns (\cref{subsubsec:evalweakscaling}), as compared to the proxy $\tilde{\mathcal{D}}$. Finally, we do a more detailed comparison between a \emph{flexible} \ac{GP}-\ac{AMG} cycle and a recursive $V(1,1)$ cycle by examining cycle complexity and convergence properties (\cref{subsec:gpamganalysis}). 
\subsubsection{Generalization to different anisotropies}
\label{subsubsec:evalanisotropy}

\begin{table}[!htbp]
    \centering
    \resizebox{\textwidth}{!}{%
    \begin{tabular}{@{}|lll||rr|rr|rr|rr|rr|rr|rr|@{}}
    \toprule
    \multicolumn{3}{|c||}{\textbf{configuration}} & \multicolumn{2}{c|}{\textbf{default}} & \multicolumn{2}{c|}{\textbf{tuned 1}} & \multicolumn{2}{c|}{\textbf{tuned 2}} & \multicolumn{2}{c|}{\textbf{tuned 3}} & \multicolumn{2}{c|}{\textbf{tuned 4}} & \multicolumn{2}{c|}{\textbf{tuned 5}} & \multicolumn{2}{c|}{\textbf{tuned 6}} \\
    \midrule
            $c_1$ & $c_2$ & $c_3$ &   T(ms) &  N &  T(ms) &  N &  T(ms) &  N &  T(ms) &  N &  T(ms) &  N &  T(ms) &  N &  T(ms) &  N \\
    \midrule
            1 & 1 & 1 &     381 & 22 &    375 & 19 &    351 & 17 &    349 & 21 &    403 & 16 &    324 & 18 &    379 & 18 \\
            1 & 1 & 1e-3 &     399 & 19 &    276 & 11 &    289 & 11 &    279 & 14 &    295 &  9 &    280 & 13 &    258 & 10 \\
            1 & 1 & 1e-5 &     398 & 19 &    275 & 11 &    288 & 11 &    278 & 14 &    298 &  9 &    280 & 13 &    256 & 10 \\
            1 & 1e-3 & 1e-3 &     365 & 21 &    190 &  9 &    194 &  9 &    219 & 13 &    \textbf{178} &  7 &    236 & 13 &    236 & 11 \\
            1 & 1e-3 & 1e-5 &     365 & 21 &    189 &  9 &    193 &  9 &    219 & 13 &    \textbf{178} &  7 &    235 & 13 &    236 & 11 \\
            1 & 1e-5 & 1e-5 &     365 & 21 &    191 &  9 &    194 &  9 &    219 & 13 &    \textbf{178} &  7 &    232 & 13 &    215 & 10 \\
    \bottomrule
    \end{tabular}%
    }
    \resizebox{\textwidth}{!}{%
    \begin{tabular}{|lll||rr|rr|rr|rr|rr|rr|rr|}
    \toprule
    \multicolumn{3}{|c||}{\textbf{configuration}} & \multicolumn{2}{c|}{\textbf{GP-22}} & \multicolumn{2}{c|}{\textbf{GP-27}} & \multicolumn{2}{c|}{\textbf{GP-28}} & \multicolumn{2}{c|}{\textbf{GP-31}} & \multicolumn{2}{c|}{\textbf{GP-33}} & \multicolumn{2}{c|}{\textbf{GP-36}} & \multicolumn{2}{c|}{\textbf{speedup}}\\
    \midrule
            $c_1$ & $c_2$ & $c_3$ &  T(ms) &  N &  T(ms) &  N &  T(ms) &  N &  T(ms) &  N &  T(ms) &  N &  T(ms) & N & $\eta_{1}$ & $\eta_{2}$ \\
    \midrule
            1 & 1 & 1 &    322 & 13 &    340 & 16 &    270 & 14 &    \textbf{208} & 12 &    245 & 16 &    259 & 21 &       1.83 &       1.56 \\
            1 & 1 & 1e-3 &    236 &  8 &    244 & 10 &    253 & 11 &    \textbf{210} & 11 &    223 & 13 &    237 & 17 &       1.90 &       1.23 \\
            1 & 1 & 1e-5 &    236 &  8 &    244 & 10 &    252 & 11 &    \textbf{210} & 11 &    222 & 13 &    238 & 17 &       1.90 &       1.22 \\
            1 & 1e-3 & 1e-3 &    228 &  9 &    233 & 11 &    237 & 12 &    214 & 12 & ,   246 & 16 &    245 & 19 &       1.71 &       0.83 \\
            1 & 1e-3 & 1e-5 &    228 &  9 &    233 & 11 &    236 & 12 &    214 & 12 &    244 & 16 &    257 & 20 &       1.71 &       0.83 \\
            1 & 1e-5 & 1e-5 &    201 &  8 &    233 & 11 &    235 & 12 &    214 & 12 &    243 & 16 &    257 & 20 &       1.82 &       0.89 \\
    \bottomrule
    \end{tabular}%
    }
        \caption{Performance of reference solvers (top) and GP solvers (bottom) for different anisotropy configurations from \cref{eq:anisotropicpoisson} with $10^6$ unknowns.}
    \label{tab:anistropyeval}
\end{table}
From $\mathcal{D}$, the problem size is fixed to $N_d=100$, and the anisotropy coefficients, $c_1,c_2,c_3$ are varied. We choose six problems, one isotropic, and the others with anisotropies of different strengths and direction, a subset of the complete evaluation set, but representative of different variants in $\mathcal{D}$ with $N_d=100$. The solve time $T$ and the number of iterations $N$ for each configuration are evaluated for the reference and \ac{GP} solvers. The speedup columns, \(\eta_1\) and \(\eta_2\), measure the speedup of the fastest \ac{GP} solver relative to the default solver and the fastest tuned solver, respectively (Table \ref{tab:anistropyeval}). \ac{GP} solvers outperform the default solver in all configurations in both the solve time and iteration count. Compared to the tuned solvers, \ac{GP} solvers are superior for isotropic problems, and for problems with weak couplings in one direction. For problems with weak couplings in multiple directions, some of the tuned solvers (tuned 1, 2, 4) are faster.  This is possibly due to the choice of the proxy $\tilde{\mathcal{D}}$ which has weak couplings only in one direction.
\subsubsection{Weak scaling}
\label{subsubsec:evalweakscaling}

\begin{table}[!htbp]
    \centering
    \resizebox{\textwidth}{!}{%
\begin{tabular}{|l||rr|rr|rr|rr|rr|rr|rr|}
\toprule
\multicolumn{1}{|c||}{} & \multicolumn{2}{c|}{\textbf{default}} & \multicolumn{2}{c|}{\textbf{tuned 1}} & \multicolumn{2}{c|}{\textbf{tuned 2}} & \multicolumn{2}{c|}{\textbf{tuned 3}} & \multicolumn{2}{c|}{\textbf{tuned 4}} & \multicolumn{2}{c|}{\textbf{tuned 5}} & \multicolumn{2}{c|}{\textbf{tuned 6}} \\
\midrule
$N_d$ &   T(ms) &   N &  T(ms) &   N &  T(ms) &   N &  T(ms) &   N &  T(ms) &   N &  T(ms) &   N &  T(ms) &   N \\

\midrule
100   &     402 &  19 &    259 &  11 &    275 &  11 &    276 &  14 &    286 &   9 &    278 &  13 &    258 &  10 \\
200    &     485 &  20 &    336 &  12 &    324 &  11 &    342 &  15 &    387 &  10 &    317 &  13 &    330 &  11 \\
300    &     530 &  21 &    380 &  13 &    368 &  12 &    352 &  15 &    448 &  11 &    350 &  14 &    406 &  13 \\
400    &    572 &  22 &    460 &  15 &    452 &  14 &    395 &  16 &    496 &  12 &    361 &  14 &    448 &  14 \\
500    &    591 &  22 &    573 &  17 &    498 &  15 &    419 &  17 &    550 &  13 &    404 &  15 &    489 &  15 \\
600    &    593 &  22 &    623 &  18 &    567 &  16 &    424 &  17 &    607 &  14 &    434 &  16 &    524 &  16 \\
700    &    611 &  22 &    610 &  18 &    575 &  16 &    487 &  19 &    666 &  15 &    491 &  18 &    567 &  17 \\
800    &    640 &  23 &    675 &  19 &    621 &  17 &    524 &  20 &    703 &  15 &    535 &  19 &    612 &  18 \\
900   &    653 &  23 &    726 &  20 &    711 &  19 &    562 &  21 &    743 &  16 &    579 &  20 &    654 &  19 \\
1000 &    657 &  23 &    772 &  21 &    720 &  19 &    627 &  22 &    837 &  17 &    590 &  21 &    698 &  20 \\
\bottomrule
\end{tabular}%
}
    \resizebox{\textwidth}{!}{%
\begin{tabular}{|l||rr|rr|rr|rr|rr|rr|rr|}
\toprule
\multicolumn{1}{|c||}{} & \multicolumn{2}{c|}{\textbf{GP-22}} & \multicolumn{2}{c|}{\textbf{GP-27}} & \multicolumn{2}{c|}{\textbf{GP-28}} & \multicolumn{2}{c|}{\textbf{GP-31}} & \multicolumn{2}{c|}{\textbf{GP-33}} & \multicolumn{2}{c|}{\textbf{GP-36}} & \multicolumn{2}{c|}{\textbf{speedup}} \\
\midrule
$N_d$ &  T(ms) &   N &  T(ms) &   N &  T(ms) &   N &  T(ms) &   N &  T(ms) &   N &  T(ms) & N & $\eta_1$ & $\eta_2$\\

\midrule
100    &    230 &   8 &    244 &  10 &    247 &  11 &    234 &  12 &    \textbf{223} &  13 &    224 &  16 &            1.80 &    1.16 \\
200   &    305 &   9 &    314 &  11 &    285 &  11 &    \textbf{275} &  12 &    279 &  14 &    297 &  18 &            1.76 &    1.15 \\
300   &    350 &  10 &    353 &  12 &    323 &  12 &    \textbf{307} &  13 &    310 &  15 &    307 &  18 &            1.72 &    1.14 \\
400   &    431 &  12 &    392 &  13 &    336 &  12 &    \textbf{318} &  13 &    320 &  15 &    337 &  19 &            1.80 &    1.14 \\
500    &    479 &  13 &    439 &  14 &    346 &  12 &    \textbf{327} &  13 &    334 &  15 &    353 &  19 &            1.81 &    1.24 \\
600    &    532 &  14 &    485 &  15 &    379 &  13 &    \textbf{361} &  14 &    368 &  16 &    371 &  19 &            1.64 &    1.18 \\
700   &    574 &  15 &    522 &  16 &    419 &  14 &    \textbf{366} &  14 &    379 &  16 &    386 &  19 &            1.67 &    1.33 \\
800   &    626 &  16 &    560 &  17 &    436 &  14 &    \textbf{378} &  14 &    382 &  16 &    405 &  20 &            1.69 &    1.39 \\
900    &    672 &  17 &    597 &  18 &    465 &  15 &    \textbf{388} &  14 &    410 &  17 &    422 &  20 &            1.68 &    1.45 \\
1000 &   1186 &  18 &    643 &  19 &    508 &  16 &    \textbf{385} &  14 &    409 &  17 &    416 &  20 &            1.70 &    1.53 \\
\bottomrule
\end{tabular}%
}
    \caption{Weak scaling results of the anisotropic Poisson problem, from the initial problem $\tilde{\mathcal{D}}$ with $10^6$ unknowns scaled to $10^9$ unknowns, for reference solvers (top) and \ac{GP} solvers (bottom).}
    \label{tab:weakscaling}
\end{table}
Now, we fix the anisotropy coefficients as set in the proxy $\tilde{\mathcal{D}}$, and vary $N_d$ to generate more problems from $\mathcal{D}$. We perform a weak scaling study with processor counts $n_p = k^3, 
\,k \in \{2,4,6,8,10,12,14,16,18,20\},$ and choose the global problem size with $N_d = 50k$.
The processor topology is constrained to a cubic domain. The scaling experiment is conducted on the Ruby compute cluster at Lawrence Livermore National Laboratory (LLNL)\footnote{https://hpc.llnl.gov/hardware/compute-platforms/ruby—decommissioned} with MPI parallelism and the results are summarized in Table \ref{tab:weakscaling}. Except \ac{GP}-22 (for $N_d=900,1000$), all \ac{GP} solvers outperform the default solver across all problem sizes. \ac{GP}-31 is the fastest solver among all GP and reference solvers, and scales the best algorithmically (12 iterations to 14 iterations). Hence, the speedup ($\eta_2$)  of \ac{GP}-31 relative to the fastest tuned solver (tuned 5) improves with an increase in the problem size. For the largest problem with $10^9$ unknowns, GP-28, -31, -33, -36, are the four fastest solvers. 
\subsubsection{Comparison of a \emph{flexible} \ac{AMG} cycle with a standard V-cycle}
\label{subsec:gpamganalysis}
G3P generates efficient AMG methods based on two fitness measures: \textit{cost per iteration} $T/N$ and \textit{convergence rate} $\rho$. Here, we derive analytical estimates for both, and compare generated \textit{flexible} AMG cycles with standard $V(1,1)$ cycles. 

\paragraph{Cost per iteration}

Let $\mathcal{W}(R)$ be the computational cost of one iteration of a method $R$ as given in \cref{eq:iterative}.
We will measure cost in work units (WUs) and define $1 \ \text{WU} := \mathcal{W}(A_L)$, that is, one application of the fine-grid operator $A_L$.
Let $\mathrm{nnz}(A)$ denote the number of nonzeros of the matrix $A$, hence, the cost of applying $A_l$ is approximated by $\mathrm{nnz}(A_l)/ \mathrm{nnz}(A_L) \ \text{WUs}.$ For the smoothers considered in this article, one relaxation sweep can be assumed to be equivalent to one application of $A_l$. Thus, the cost of one multigrid cycle $M$ is approximated by

\[
\mathcal{W}(M)
\approx
\sum_{l=1}^{L}
\nu_l
\frac{\mathrm{nnz}(A_l)}{\mathrm{nnz}(A_L)} \text{ }\mathrm{WUs}, 
\]
where, $\nu_l$ denotes the number of smoothing steps on level $l$.
This can be further simplified for our proxy problem $\tilde{D}$, solved using BoomerAMG with the setup parameters stated in \cref{sec:designconstraints}, as

\[
\mathcal{W}(M)
\approx
\nu_1(1.2\!\times\!10^{-5})
+ \nu_2(9.5\!\times\!10^{-5})
+ \cdots
+ \nu_8(1.9)
+ \nu_9(1.0)\text{ }\mathrm{WUs}.
\]

Substituting the smoothing parameters gives

\begin{equation}
    \mathcal{W}(M^{\text{GP-31}}) \approx 6.7\text{ }\mathrm{WUs},
\qquad
\mathcal{W}(M^{V}) \approx 8.5\text{ }\mathrm{WUs}.
\label{eq:costgp31}
\end{equation}

\begin{figure}[!htbp]
    \centering
    \includegraphics[width=1\linewidth]{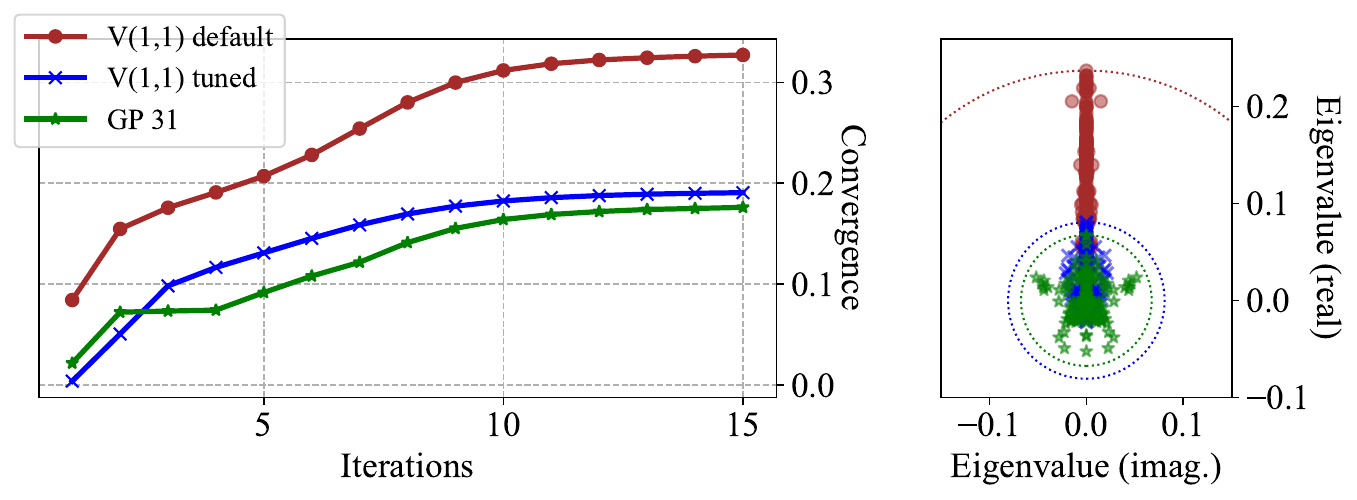}
    \caption{Convergence plots of \ac{GP}-31, default $V(1,1)$, and `tuned-1' $V(1,1)$ \ac{AMG} solver on problem $\tilde{\mathcal{D}}$ (left), and the eigenvalue distribution of the respective iteration matrices on a smaller $10^3$ variant (right).}
    \label{fig:convcomparison}
\end{figure}

The cost reduction in GP-31 is obtained by eliminating smoothing on the most expensive level $l=8$ with $\nu_8=0$, and reallocating it to the finest level with $\nu_9=4$ (see \cref{fig:GP31}), whereas the standard $V(1,1)$ cycle uses
$\nu_8=\nu_9=2$ maintaining a recursive structure.
\paragraph{Convergence rate}
The eigenvalues of the iteration matrices for GP-31, and standard $V(1,1)$ cycles with the tuned (`tuned-1') and default configurations are plotted on a smaller problem with $10^3$ unknowns, but problem properties identical to $\tilde{D}$. We measure the spectral radius as a measure of the \emph{asymptotic convergence}, that is, the worst error reduction over many iterations. The right side of \cref{fig:convcomparison} indicates that GP-31 has better asymptotic convergence properties compared to standard $V(1,1)$ cycles, despite a lower \emph{cost per iteration}, as shown in \cref{eq:costgp31}. This aligns qualitatively with the numerical results on $\tilde{\mathcal{D}}$ (\cref{fig:convcomparison}, left). 

%% file: siamart_220329/contents/results.tex
\section{Automated design of an \ac{AMG} preconditioner for a multiphysics problem}
Next, we employ AMG within a problem from the ARES multiphysics code at LLNL \cite{McAbee2001, Bender2021, vcs5-7ymc}. Solving the radiation diffusion equation within ARES is often computationally expensive, and its general form in relation to \cref{eq:continuousform} can be written as

\begin{equation}
    \label{eq:radiationdiffusioncontinuous}
    E_t + (c\,d\,\kappa \,T)E - \nabla \cdot (D(E) \nabla E) = f \quad \text{in $\Omega$ $\times$ [$0,T$]}.
\end{equation}

Here, $E$ is the radiation energy per unit volume, $c$ is the speed of light, $d$ is the density, $\kappa$ is the opacity, $T$ is the matter temperature, and $D$ is the diffusion coefficient. This general form is non-linear due to the dependency of the diffusion coefficient $D(E)$ on the radiation energy density $E$ from flux limiting. After making a set of spatial discretization choices and treating $D$ as fixed, \cref{eq:radiationdiffusioncontinuous} becomes linear, and we arrive at a semi-discrete form over $N$ control-volume cells as

\begin{equation}
    M_h E(t) + K_hE + C_hE =F_h \quad \text{in $\Omega_h$ $\times$ [$0,T$]},
\end{equation}
with $M_h := \mathrm{diag}(V_1,V_2,\dots,V_N)$, $K_hE \approx \nabla_h\cdot(D\nabla_hE)$, and $C_h := \mathrm{diag}\left((c\,d_1\,\kappa_1 T_1),\dots,(c\,d_N\,\kappa_N T_N) \right)$, where $V_i$ is the control-volume cell measure associated with degree of freedom (DOF) $i$. By employing the backward Euler method as the implicit time integration scheme, one solves a sparse linear system at each step, of the form

\begin{equation}
\label{eq:radiationlinear}
    \underbrace{\left(M_h + \Delta t(K_h + C_h)\right)}_A \underbrace{E^{(n+1)}}_\mathbf{x} = \underbrace{M_hE^{(n)} + \Delta tF_h^{(n)}}_\mathbf{b}.
\end{equation}

\begin{figure}
    \centering
    \begin{subfigure}{0.48\textwidth} 
        \centering
        \includegraphics[width=\textwidth]{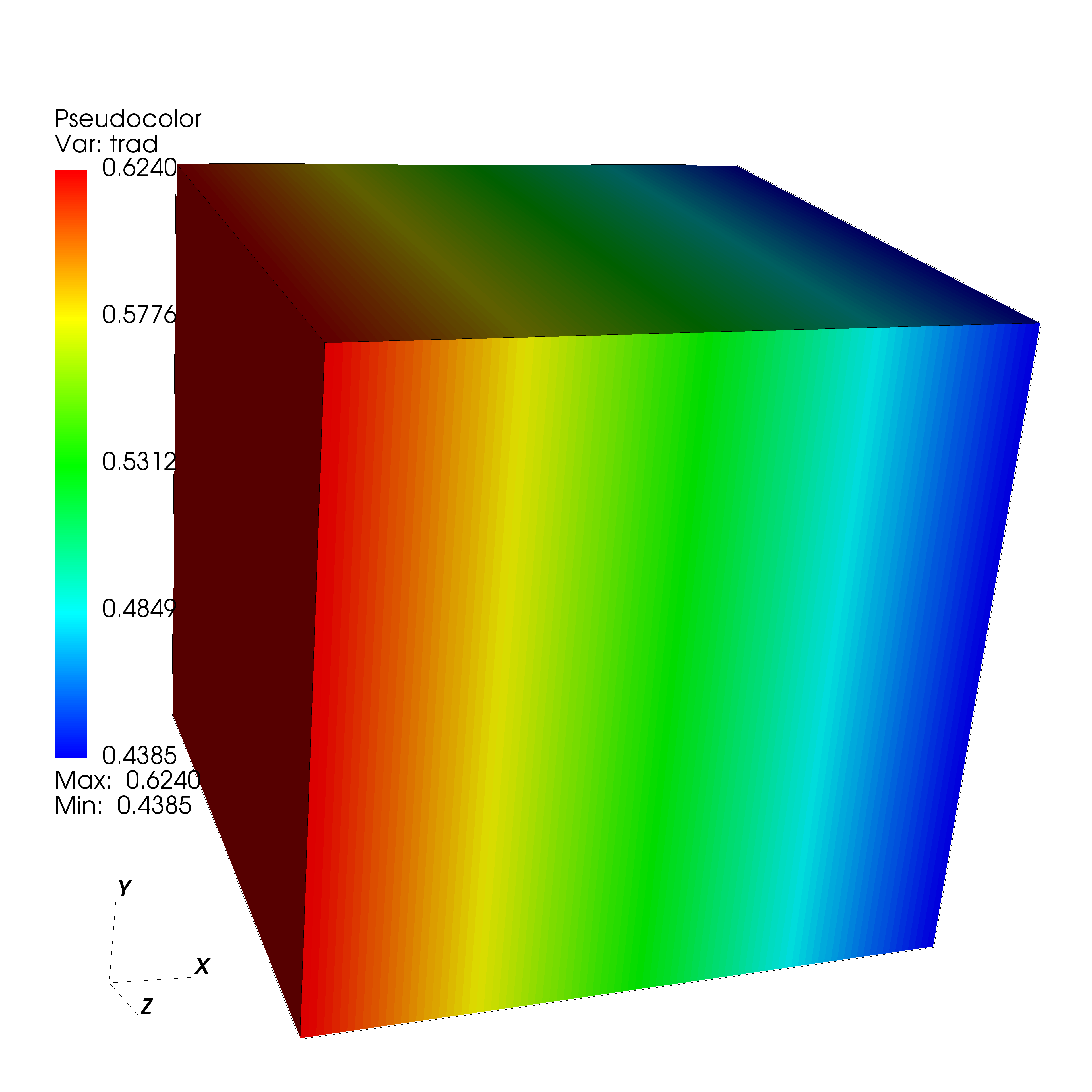} 
        \caption{Initial radiation temperature of zrad3D problem (in units of KeV).}
        \label{fig:sub_a}
    \end{subfigure}
    \hfill 
    \begin{subfigure}{0.48\textwidth} 
        \centering
        \includegraphics[width=\textwidth]{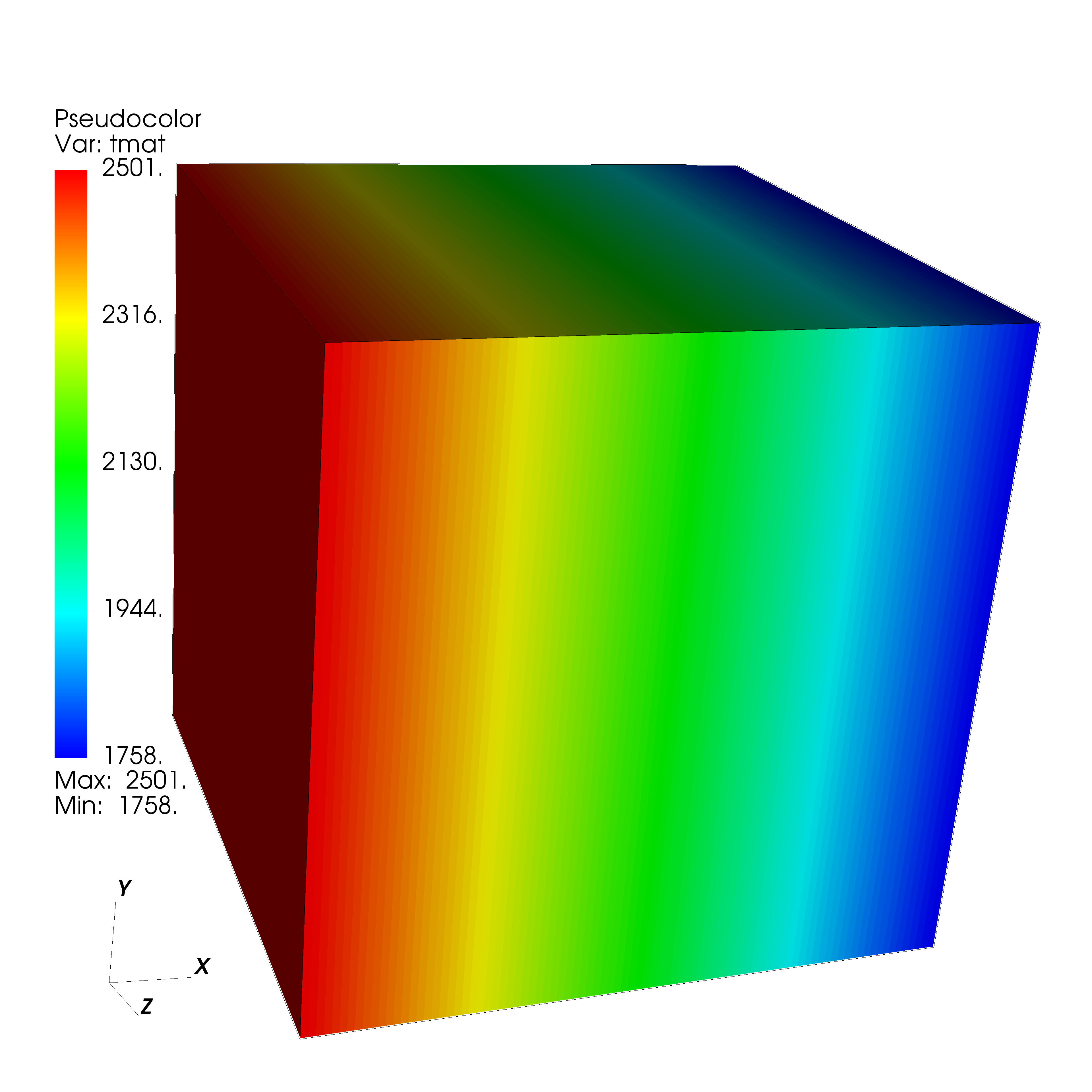} 
        \caption{Initial matter temperature of zrad3D problem (in units of KeV).}
        \label{fig:sub_b}
    \end{subfigure}
    \caption{Initial conditions of the \emph{zrad3D} problem. A source is applied to the left side of the block as it is pictured above. As the problem evolves, radiation flows through the block, and the matter and radiation fields come into equilibrium with each other.}
    \label{fig:dual_figure}
\end{figure}

We now employ this discretized system on an application problem that provides a stress test for the underlying linear solver, a benchmark dubbed the \emph{zrad3D} problem.  The equations being solved in this problem are the radiation diffusion and electron conduction equations, the latter discretized in a similar fashion to the radiation diffusion equation. The problem involves radiation flow through a 3D block with $N_d^3$ unknowns. Although topologically simple, this problem is set up with a large difference between radiation and matter temperatures, and with an appreciable temperature gradient across the block (\cref{fig:dual_figure}). This results in SPD matrices with strong connections between unknowns corresponding to off-diagonal matrix entries, resulting in ill-conditioned systems that can cause a diagonally scaled conjugate gradient (CG) method to converge very slowly, even for a relatively small problem size. The solver and physics packages stress tested in the \emph{zrad3D} problem are identically deployed in inertial confinement fusion (ICF) simulations, therefore, results regarding the efficacy of an AMG preconditioner in \emph{zrad3D} may help inform the solver strategy and performance of production ICF calculations. Hence, we design an efficient AMG preconditioner for \emph{zrad3D} using G3P for linear systems $\mathcal D = \left\{\{ A^{(k)} \}_{k=1}^{40} \;
\middle|\;
A^{(k)} \in \mathbb{R}^{N_d^3 \times N_d^3},
\quad
N_d \in \{100,101, \dots,400\}\right\}$ from the first 40 time steps of the simulation, alternating between \cref{eq:radiationlinear} for radiation diffusion when \emph{k is odd} and its corresponding discretized form for electron conduction when \emph{k is even}. We choose a right-hand side (RHS) with random entries, to avoid the special case where the RHS lies in a low-dimensional subspace of the eigenmodes, potentially leading to artificial early Krylov convergence. The preconditioner is applied once per CG iteration, and we solve to a relative residual tolerance of $\epsilon=10^{-6}$.  A proxy $\tilde{\mathcal{D}}=\{A^{(1)} \;|\; A^{(1)} \in \mathbb{R}^{100 \times 100}\}$, being the most ill-conditioned system in $\mathcal{D}$ for the smallest problem $N_d=100$, is used for fitness computation in G3P and solved across $n_p=8$ MPI processes. We measure \emph{time} $T$ for the \emph{preconditioned CG solve} (excluding AMG setup time), \emph{number of CG iterations} $N$, preconditioned CG convergence rate $\rho$ to solve $\tilde{\mathcal{D}}$; for each individual AMG preconditioner $x$, and assign its fitness $\mathbf{f_x} = [T/N, \rho]^T$.

\label{sec:amgpredconditioner}

\subsection{Reference preconditioners}
Similar to choosing competitive reference solvers for the anistropic Poisson problem as seen earlier in \cref{subsec:refamgsolvers}, \ac{AMG} preconditioners are tuned for the proxy \emph{zrad3D} problem $\tilde{\mathcal{D}}$. A $V(1,1)$ \ac{AMG} preconditioner with a $l_1$ Jacobi smoother is found to be the fastest \ac{PCG} solver. This \ac{PCG} solver will be henceforth called `tuned'; and the \ac{PCG} solver with the default \ac{AMG} preconditioner (listed as default in \cref{tab:refsolvers}) is called `default'. 

\subsection{Performance Evaluation and Generalizability}
\begin{figure}[!htbp]
    \centering
    \begin{minipage}{0.69\linewidth}
        \centering
        \includegraphics[width=\linewidth]{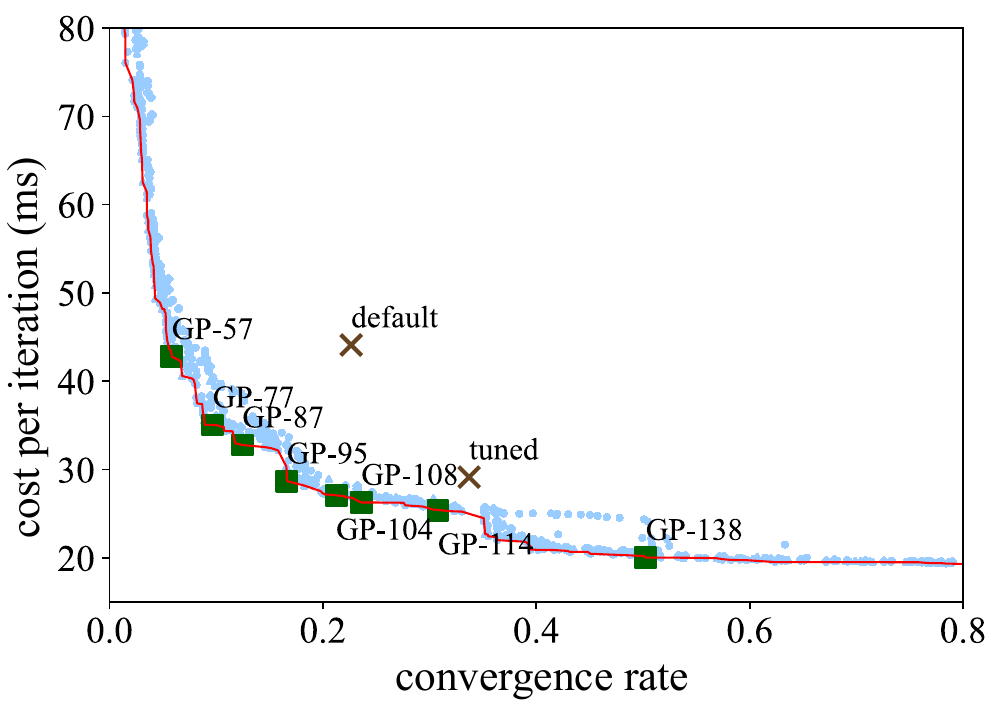}
    \end{minipage}
    \hfill
    \begin{minipage}{0.30\linewidth}
        \centering
        \begin{tabular}{|l||rr|}
        \hline
        Solver & T(ms) & N \\
        \hline
        GP-57 & 214 & 5 \\
        GP-77 & 210 & 6 \\
        GP-87 & 230 & 7 \\
        GP-95 & 230 & 8 \\
        GP-104 & 244 & 9 \\
        GP-108 & 263 & 10 \\
        GP-114 & 305 & 12 \\
        GP-138 & 421 & 21 \\
        tuned & 387 & 13 \\
        default & 448 & 10 \\
        \hline
        \end{tabular}
    \end{minipage}
    \caption{(Left) Final population of evolved PCG programs, shown as blue dots, with the resulting Pareto front highlighted in red line. (Right) Comparison of solve time and CG iterations between the selected GP-PCG programs and the reference PCG configurations for the proxy $\tilde{\mathcal{D}}$.}
    \label{fig:pareto_and_table}
\end{figure}
The \ac{PCG} programs with the \ac{CFG} generated \ac{AMG} preconditioners are evolved for $100$ generations, resulting in a final population as shown in \cref{fig:pareto_and_table}. $8$ GP-PCG programs, denoted by green squares, are picked from different regions of the final Pareto front. The default and the tuned \ac{PCG} programs are represented by a brown cross, and their positions lie outside the Pareto front, implying their suboptimal performance compared to the generated GP-PCG programs. The performance of the picked GP-PCG programs with the reference programs, on the proxy problem is also listed in \cref{fig:pareto_and_table} (on the right).  The fastest GP-PCG programs achieve speedups of approximately $2.1\times$ and $1.8\times$ compared to the default and tuned \ac{PCG} programs, respectively. 
\begin{figure}[!htbp]
    \centering
    \includegraphics[width=0.9\linewidth]{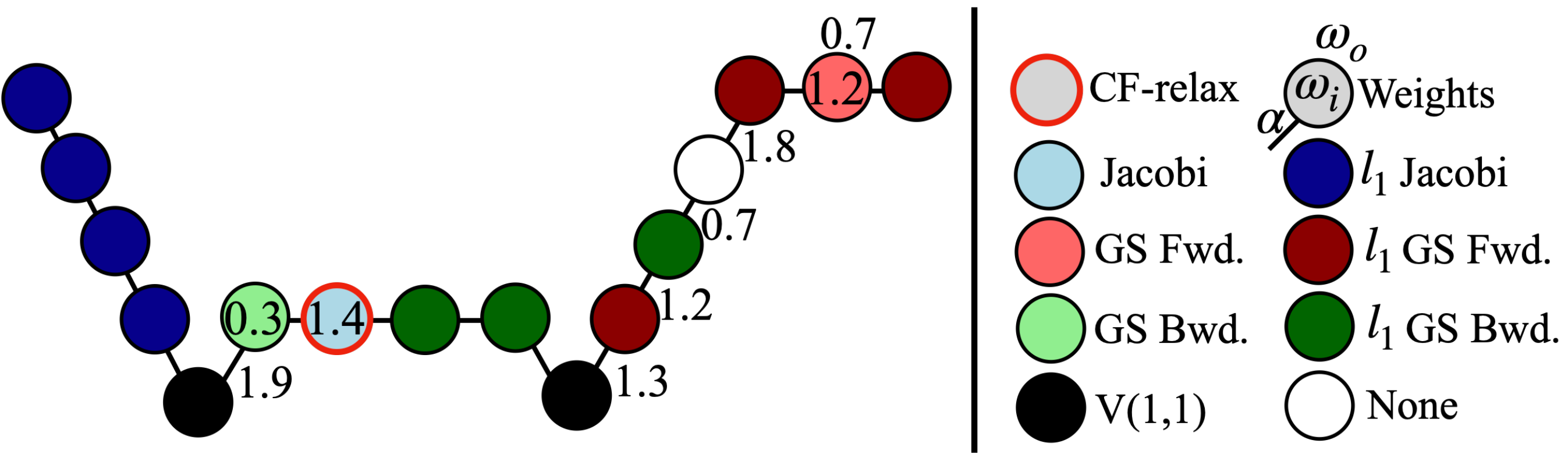}
    \caption{Visual representation of the GP-57 AMG preconditioner cycle.}
    \label{fig:GP57}
\end{figure}
Furthermore, these programs are evaluated for generalizability to the problem set $\mathcal{D}$, in \cref{subsec:arestimestepping,subsec:aresweakscaling}. We also show that our GP solvers with novel AMG \emph{flexible} cycling can be used with other wrappers in \emph{hypre} (\cref{subsec:areshybrid}), demonstrating its compatibility as a valid building block in the larger software framework of \emph{hypre}.

\subsubsection{Generalization to different time-steps}
\label{subsubsec:gentimesteps}
\begin{figure}[!htbp]
    \centering
    \includegraphics[width=1\linewidth]{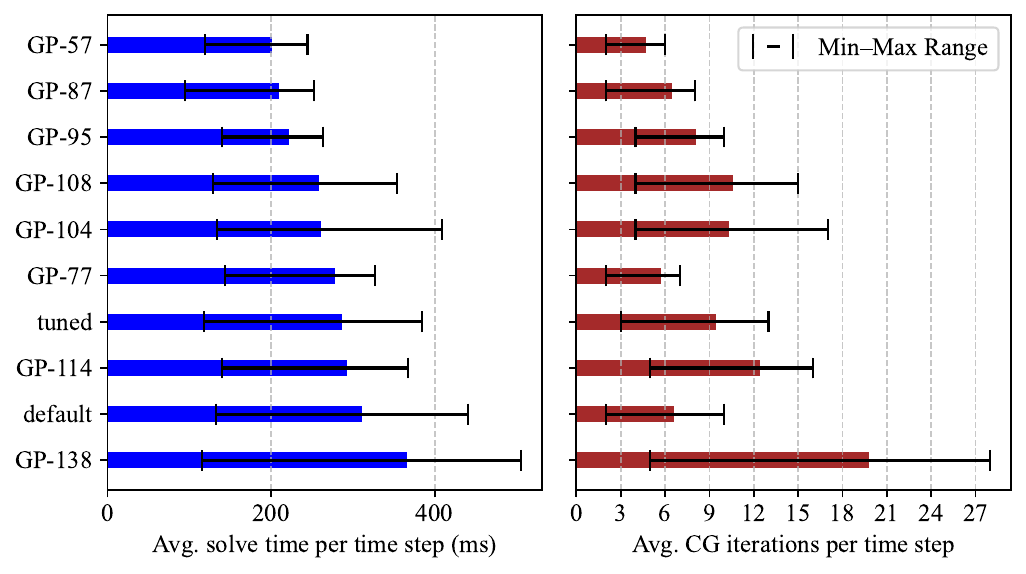}
    \caption{Performance of GP-PCG solvers compared to reference solvers, measured over the first 40 time steps of the \emph{zrad3D} problem with $10^6$ DOFs.}
    \label{fig:gpares_comparison}
\end{figure}

The selected GP-PCG programs, generated based on the proxy $\tilde{\mathcal{D}}$, are now evaluated for matrices from the first 40 time steps of the \emph{zrad3D} problem with $10^6$ DOFs. The solve time and number of CG iterations are averaged over these 40 time steps for GP-PCG solvers and the reference solvers (\cref{fig:gpares_comparison}). The solvers are ordered from fastest to slowest. As shown, 6 out of the 8 selected GP-PCG programs outperform both reference solvers, while 7 outperform the default solver. On average, the fastest GP-PCG program, \textit{GP-57} (\cref{fig:GP57}), achieves a speedup of approximately $1.54\times$ and $1.42\times$ over the default and tuned solvers, respectively. Also, GP-57 shows a narrower distribution of solve times and CG iteration counts throughout the simulation, compared to the reference solvers, highlighting its robustness, despite being optimized only on a single linear system in $\tilde{\mathcal{D}}$.
\label{subsec:arestimestepping}
\subsubsection{Weak scaling}
\label{subsec:aresweakscaling}
We now assess the performance of GP-57, tuned, and default solvers on the full problem set $\mathcal{D}$, by performing a weak scaling study with $N_d \in \{100,200,400\}$ evaluated over the first 40 time steps of the \emph{zrad3D} problem. The weak scaling study is done on a different CPU platform, \textit{RZHound}\footnote{\url{https://hpc.llnl.gov/hardware/compute-platforms/rzhound}} at LLNL, demonstrating solver portability. The local problem is fixed to 25,000 DOFs per MPI process, and uniformly refined in all three directions. The weak scalability results are summarized in \cref{tab:weakscaling_aresGP} with their corresponding total solve times and total CG iterations over the first $40$ time steps. GP-57 achieves faster solve times with fewer CG iterations compared to both default and tuned solvers. Also, the net speedup of GP-57, $\eta_1$ and $\eta_2$, relative to the default and tuned solvers, respectively, remains consistent and generalizes well to larger problem sizes. 
\begin{table}[!htbp]
\centering

\begin{tabular}{|c || r r | r r | r r | r r|}
\hline
 & \multicolumn{2}{c|}{\textbf{default}} & \multicolumn{2}{c|}{\textbf{tuned}} & \multicolumn{2}{c|}{\textbf{GP-57}} & \multicolumn{2}{c|}{\textbf{speedup}}  \\
\hline
DOFs        & T(s) & N & T(s) & N & T(s) & N & $\eta_1$ & $\eta_2$ \\
\hline
1M   & 9.0  & 267 & 8.8  & 373 & \textbf{5.4} & 172 & 1.66 & 1.63 \\
8M   & 12.4 & 323 & 11.8 & 433 & \textbf{7.7} & 210 & 1.60 & 1.53 \\
64M  & 15.2 & 387 & 14.7 & 519 & \textbf{9.2} & 240 & 1.66 & 1.60 \\
\hline
\end{tabular}
\caption{Weak scaling results of default, tuned and GP preconditioned CG solvers for the \emph{zrad3D} problem evaluated over the first $40$ time steps on the \textit{RZHound} system.}
\label{tab:weakscaling_aresGP}
\end{table}

\subsubsection{Compatibility of GP solvers with other interfaces in \textit{hypre}}
\label{sec:hybridsolvers}
\begin{figure}[!htbp]
    \centering
    \includegraphics[width=1\linewidth]{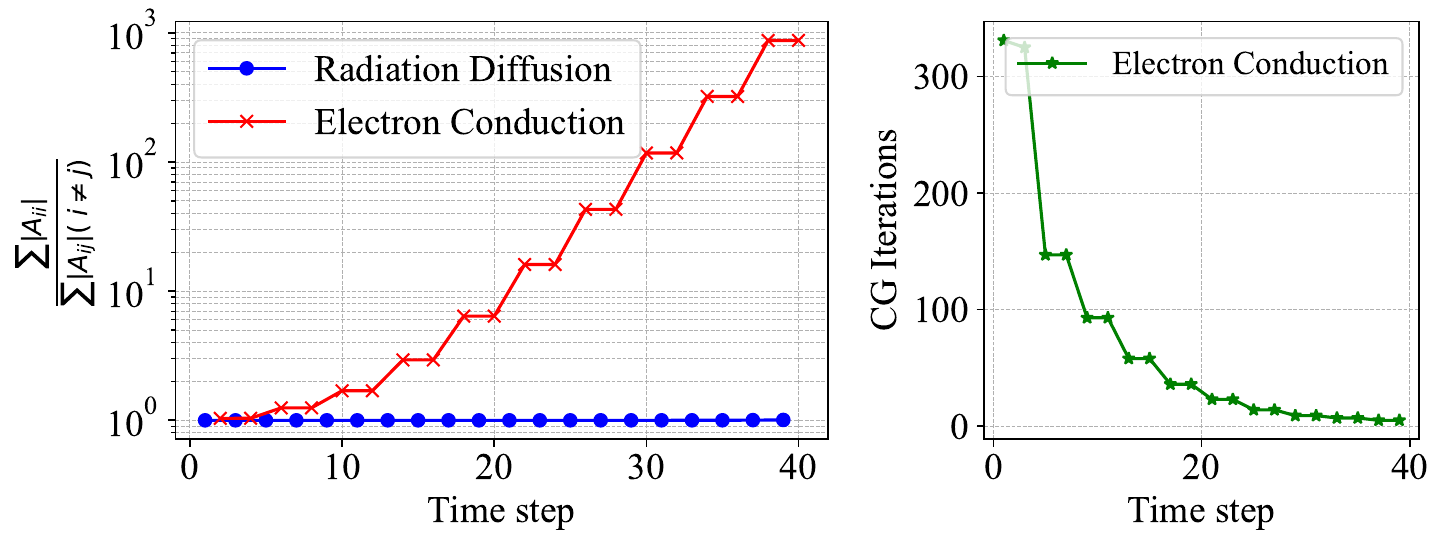}
    \caption{(Left) The relative contribution of diagonal versus off-diagonal elements in the system matrix throughout the simulation. (Right) CG convergence of the electron conduction problem with a simple diagonal preconditioner.}
    \label{fig:diagonoldominance}
\end{figure}

\begin{figure}[!htbp]
    \centering
    \includegraphics[width=1\linewidth]{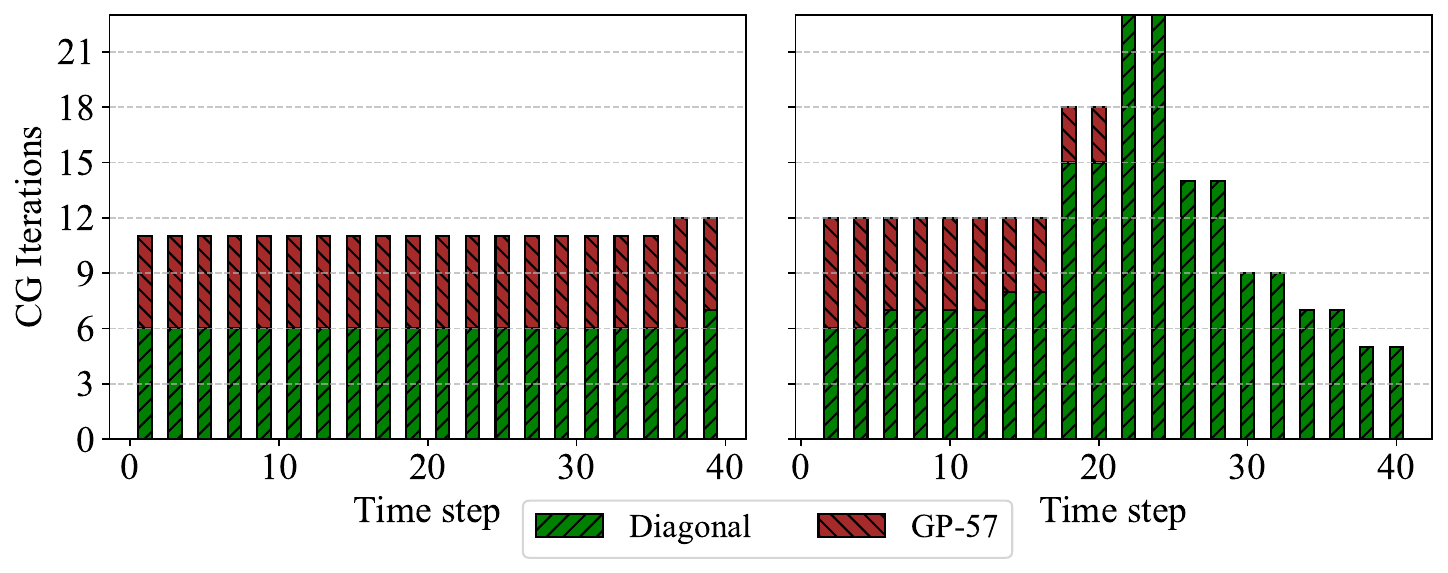}
    \caption{CG convergence on the radiation diffusion (left) and electron conduction (right) systems across the simulation, using the \textit{hybrid solver} interface with the GP-57 preconditioner, and a convergence threshold set to $0.65$.}
    \label{fig:gphybrid_ares}
\end{figure}

To assess how strongly the system matrix $A$ with entries $a_{ij}$ is dominated by its diagonal, we define a quantity, 
\[
\eta(A):=\frac{\displaystyle \sum_{i=1}^{N_d^3} |a_{ii}|}
{\displaystyle \sum_{i=1}^{N_d^3} \sum_{\substack{j=1 \\ j\neq i}}^{N_d^3} |a_{ij}|},
\]
which measures the relative contribution of the diagonal entries compared to the off-diagonals. $\eta(A^{(k)})$, when \emph{k is even}, that is, for the electron conduction equation, increases during the simulation (\cref{fig:diagonoldominance}, left) and hence, using a simple diagonal preconditioner is sufficient for a time step k when $\eta(A^{(k)})$ is sufficiently large, (\cref{fig:diagonoldominance}, right). This helps avoid expensive AMG setup costs. However, an AMG is still required for solving the radiation diffusion problem, and for the earlier part of the electron conduction problem. The \textit{hybrid solver} interface in hypre\footnote{https://hypre.readthedocs.io/en/latest/solvers-hybrid.html} is designed for such cases by automatically switching from a simple diagonal scaling to a more expensive AMG preconditioner based on a user-defined convergence threshold, that is, if the convergence rate of CG with diagonal scaling degrades beyond this prescribed threshold, the solver sets up an AMG preconditioner; but otherwise continues using a diagonal preconditioner. Since we use GP guided by a \ac{CFG} that embeds the algorithmic components of BoomerAMG (\cref{fig:bnfboomeramg}), the compatibility of the generated GP solvers with other interfaces in hypre is preserved. Hence, we can directly plug in our generated GP solvers to the  \emph{hybrid solver} interface for our simulation, as shown, in \cref{fig:gphybrid_ares}. GP-57 is used to accelerate CG convergence beyond simple diagonal scaling, hence avoiding excessively large iteration counts. For the electron conduction problem (\cref{fig:gphybrid_ares}, right), the use of AMG preconditioners is reduced as the system’s conditioning improves, and beyond the $22^{nd}$ time step, AMG is no longer required, thereby saving expensive setup costs.
\label{subsec:areshybrid}

%% file: siamart_220329/contents/conclusion.tex
\section{Cost of AMG evolution}
\label{sec:cost}
The generation of AMG methods for problems in \cref{sec:amgsolver} and \cref{sec:amgpredconditioner} requires approximately \(29\,\mathrm{min}\) and \(55\,\mathrm{min}\), respectively,  when distributed over 32 nodes across 64 MPI processes. This corresponds to a total computational cost of \(15.5\) and \(29.3\) node-hours, respectively. Evolving a single generation of AMG solvers in \cref{sec:amgsolver} takes \(13\)--\(24\,\mathrm{s}\), while evolving a single generation of AMG preconditioners for the time-stepping simulation in \cref{sec:amgpredconditioner} requires \(23\)--\(49\,\mathrm{s}\). It is worth reiterating that our approach works entirely without prior data or labeled examples of well-performing solvers, in contrast to many supervised learning techniques that depend on well-curated training datasets. Hence, in essence, the intent behind our approach is to offload the manual effort and cost associated with designing application-specific solvers onto computational resources.

\section{Conclusions}
\label{sec:conclusions}
We have demonstrated an automated approach to design AMG methods that can be integrated non-intrusively with existing solver frameworks such as \emph{hypre}. By formulating a CFG that encodes the algorithmic and syntactic constraints of \emph{BoomerAMG}, the AMG implementation in \emph{hypre}, we generate novel \emph{flexible} multigrid cycles for a fixed AMG setup using G3P. These \emph{flexible} AMG methods are generated for a stationary test problem (\cref{sec:amgsolver}) and a time-stepping simulation code (\cref{sec:amgpredconditioner}). In both cases, our approach generates \emph{flexible} AMG methods that outperform not only the default AMG method but also hand-tuned AMG V-cycles, with faster solve times and improved convergence rate. Furthermore, these methods generalize well, maintaining their superior performance across different variants of the stationary problem (\cref{subsubsec:evalanisotropy}), under system matrix perturbations during time-stepping (\cref{subsec:arestimestepping}), when scaled to larger problem sizes (\cref{subsubsec:evalweakscaling,subsec:aresweakscaling}), and also when embedded within other \emph{hypre} interfaces (\cref{sec:hybridsolvers}). The problems chosen are diffusion-dominated, where AMG is well understood and performs strongly, and this choice was deliberate for rigorous benchmarking. Our intent was to test whether a set of automatically evolved AMG programs, starting from a random initial population, could remain competitive with highly optimized multigrid baselines. Remarkably, not only do we remain competitive, but we are also able to surpass them, possibly attributable to the added cycling flexibility, a feature that is intractable to tune by hand. 

A natural extension of our approach is to customize the grammar to include the setup parameters, thereby generating complete AMG methods. Also, there is potential to leverage our framework to design AMG methods for modern GPU clusters since non-standard cycle types like $\kappa$-cycles have already been proven to be beneficial for such platforms \cite{Avnat2023-nr}. Adding more AMG ingredients can bring further improvements on GPUs, by finding algorithms that effectively exploit multiple precisions \cite{YuHsiang2024, Vacek2024}, and incur low communication scaling costs \cite{mitchell2021parallel-3c0}. Beyond classical multigrid approaches, the grammar could be extended with additional algorithmic building blocks to design composite solvers for coupled systems such as thermo elasticity \cite{parthasarathy2025proceedings-016, bevilacqua2025large-scale-162} or Stokes problems \cite{kohl2022textbook-5f1, voronin2025monolithic-9af, kruse2021parallel-6b0}. Since our framework is agnostic to the underlying PDE technology, it could also be tailored to design hybrid approaches that combine neural networks with classical mesh-based methods \cite{SolverInTheLoop2020,zhang2024blending-492,bouziani2024differentiable-651}, by training the networks themselves using \emph{neuroevolution}\footnote{An evolutionary technique that evolves both the topology and weights of neural networks (https://neuroevolutionbook.com).}\cite{risi-et-al:book2025,assuno2019denser-3a1}. Furthermore, there remains considerable scope to improve the quality and cost of the solver generation pipeline, for instance, by learning a better initial population using large-language models for faster convergence \cite{NeuralguidedGP}, and by limiting the evaluation cost for more sophisticated fitness functions (for example, solving multiple right-hand sides per candidate) through statistical techniques like \emph{racing procedures} \cite{racingalgs, lpez-ibez2016irace-03e}. Since our approach is predominantly an offline procedure exploring a very large search space, future work could also focus on coupling it with online auto-tuning frameworks such as \emph{GPTune} \cite{liu2021gptune-0e3}, by identifying and transferring a reduced set of critical parameters to such online frameworks, for highly sensitive and non-stationary simulations.

%% file: siamart_220329/contents/acknowledgements.tex
\section*{Acknowledgments}

This work was performed under the auspices of the U.S. Department of Energy by Lawrence Livermore National Laboratory under Contract DE-AC52-07NA27344. LLNL release number: LLNL-JRNL-2013976-DRAFT.
The United States Government retains, and the publisher, by accepting the article for publication, acknowledges that the United States Government retains a non-exclusive, paid-up, irrevocable, world-wide license to publish or reproduce the published form of this manuscript, or allow others to do so, for United States Government purposes.

The research of the 5th author was co-funded by the financial support of the European Union under
the REFRESH  --Research Excellence
For Region Sustainability and High-Tech Industries-- 
Project No.~CZ.10.03.01/00/22\_003/0000048
via the Operational Programme Just Transition.